\documentclass[a4paper,10pt]{article}
\usepackage[affil-it]{authblk}
\usepackage{amssymb,latexsym,amsmath,amsfonts}
\usepackage{graphicx}
\usepackage{subfigure}
\usepackage{natbib}
\usepackage{color}
\usepackage[english]{babel}

\newcommand{\be}{\begin{equation}}
\newcommand{\ee}{\end{equation}}
\newcommand{\bea}{\begin{eqnarray}}
\newcommand{\eea}{\end{eqnarray}}
\newcommand{\beas}{\begin{eqnarray*}}
\newcommand{\eeas}{\end{eqnarray*}}

\begin{document}

\title{Analysis of complex singularities in high-Reynolds-number Navier-Stokes solutions}

\author{F. Gargano,  
\thanks{email: \texttt{gargano@math.unipa.it}}}

\author{M. Sammartino
\thanks{email: \texttt{marco@math.unipa.it}}}

\author{V. Sciacca
\thanks{email: \texttt{sciacca@math.unipa.it}}}
\affil{Department of Mathematics, University of Palermo, Italy}

\author{K.W. Cassel 
\thanks{email: \texttt{cassel@iit.edu}}}
\affil{Department of Mechanical, Materials, and Aerospace Engineering,
Illinois Institute of Technology, Chicago, USA}

\date{}

%
%


\maketitle

\begin{abstract}
Numerical solutions of the laminar Prandtl boundary-layer and Navier-Stokes equations are considered for
the case of the two-dimensional uniform flow past an impulsively-started circular cylinder. 
We show how Prandtl's solution develops a finite time separation singularity.
On the other hand Navier-Stokes solution is characterized by the presence of two kinds of viscous-inviscid interactions that can be detected
by the analysis of the enstrophy and of the pressure gradient on the wall.
Moreover we  apply the complex singularity tracking method to  Prandtl and Navier-Stokes 
solutions and analyze the previous interactions from a different perspective.
\end{abstract}


\section{Introduction}

The study of the behavior of a high-Reynolds-number fluid interacting with a
solid boundary is a central problem
both in the mathematical analysis of fluid dynamics as well as
in many practical applications. This is in large part due to imposition of the no-slip boundary condition at solid surfaces 
that creates a strong gradient in the normal direction with a large amount of vorticity generated. 
 In particular, in the limit as Reynolds number $Re$ goes to infinity, the convergence of the Navier-Stokes solution to the Euler solution
close to the boundary fails, giving rise to the need for Prandtl's boundary layer along solid surfaces.
In the presence of an adverse pressure gradient, the thin boundary layers near solid surfaces are susceptible to separation. 
In classical non-interactive boundary-layer theory, which is the framework of our investigation, as the pressure gradient is imposed by the outer flow,
no viscous-inviscid interaction is permitted between the  boundary layer and the Euler flow, and the unsteady boundary-layer equations may break 
down in the form of a separation singularity within finite time.  
This was first suggested by \cite{Bla08} and later shown numerically by \cite{vDS80} in the case of the
impulsively-started circular cylinder.  Similar behavior was later shown for other initial conditions (\cite{PSW91a,DW84,Cas00}).
In all of these cases, the adverse pressure gradient causes the boundary layer to thicken rapidly in a very narrow streamwise region as the separation singularity is approached.  
It must be pointed out, however, that the imposed adverse pressure gradient is not
the only condition that eventually leads to singularity formation in 
Prandtl's solutions.  In \cite{EE97}, the authors introduced a set of initial data for
Prandtl's equation in which no adverse pressure gradient is imposed, but a
singularity forms in a finite time.
The viscous-inviscid interactions in Navier-Stokes solutions at finite Reynolds numbers behave in a different manner from that observed in the classical non-interactive 
Prandtl boundary-layer solution.
A review of the various stages in the unsteady separation
process in different Reynolds-number regimes can be found in \cite{Cas00,OC02} and \cite{GSS11}, where
the boundary layer induced by a thick-core vortex and a rectilinear vortex have been numerically simulated for different Reynolds numbers.
The occurrence of two distinct viscous-inviscid
interactions acting over different length scales has been detected.
The first interaction, called {\it large-scale interaction}, is found to occur for
all finite Reynolds numbers, and it acts over a scale that is comparable with the
characteristic length and velocity typical of the problem considered.
The large-scale viscous-inviscid interaction is the precursor to a
{\it small-scale interaction} that develops only for moderate to high Reynolds numbers 
(generally $Re\geq O(10^4)$), and this small-scale interaction is related to
the formation of a local minimum in the streamwise pressure gradient on the
boundary.  The small-scale interaction stage is marked by especially large gradients in vorticity and other flow quantities,
a splitting of the primary recirculation region, and a very narrow ejection from within the boundary layer. 
This is followed by formation of small-scale vortical structures and a large amount of vorticity production on the boundary that in turn leads to the growth of enstrophy.
In the present study, these same features will be identified in the impulsively-started flow around a circular cylinder.
Both large-scale and small-scale interactions at finite Reynolds numbers begin quite early with
respect to the first viscous-inviscid interaction that occurs in the Prandtl
boundary-layer solution, which is responsible for the ultimate failure of Prandtl's
equations to give accurate approximations of the Navier-Stokes solutions for finite Reynolds numbers.
In \cite{Cas00} and \cite{OC02}, it has been conjectured that as $Re\rightarrow\infty$, both
interactions merge and behave in the same way as the mechanisms leading to the van Dommelen and Shen (VDS) singularity.
It will be shown here that this conjecture is also supported by the present results, particularly with the aid of the complex singularity analysis performed 
on the Navier-Stokes and Prandtl's solutions.
This analysis is carried out using the wall shear and streamwise velocity component of the Navier-Stokes
solutions.
An investigation of the singularity formation of Prandtl's boundary-layer equations in the case
of the uniform flow past an impulsively started circular cylinder has been performed in \cite{DLSS06,GSS09}.
The complex singularity analysis has been applied on the streamwise velocity component of
Prandtl's solution, and it has been shown that a complex singularity does not appear \textit{out of the blue},
but it stays  in the complex domain and hits the real axis in a finite time.
This singularity is classified
as a cubic-root singularity.
In the present work, this methodology will be utilized along with the Borel-P\'olya-Hoeven (BPH) method in \cite{PF07}
and the Pad\'{e} approximation in \cite{BGM96} to characterize and track the positions of the singularities in the
complex plane.

%
The separation process and the validity of the boundary-layer
approximation also will be investigated and explained through study of the
complex singularities in the streamwise velocity component of Navier-Stokes solutions. This study is
performed by extending the singularity-tracking method to the two-dimensional
case.  In doing so, it will be determined that the rates of exponential decay $\delta_{NS}$ and algebraic decay $\alpha_{NS}$ 
of the shell-summed Fourier amplitudes determine the width of the analyticity strip of the solution and the
characterization of the complex singularity nearest to the real axis.
A similar investigation has been performed in \cite{GSS09} for the streamwise velocity component of Prandtl's solution, 
and the results obtained here for the Navier-Stokes equation will be compared with those results. 
%
%
As mentioned earlier, similar viscous-inviscid interactions have been  
observed in various settings. 
For example, \cite{BW02} simulated the unsteady separation process induced by symmetric
counter-rotating streamwise vortices for high Reynolds numbers. 
They found a similar eruptive behavior in
the boundary-layer flow; moreover, the possible presence of a physical instability has been revealed by the formation of high-frequency oscillations
in the solution. The instability is postulated to be a Rayleigh instability, as it forms after an
inflection point in the streamwise velocity.
However, it was  shown in \cite{OC05} that the instabilities of the Navier-Stokes solution in \cite{BW02} disappear with a finer computational grid.  
The possibility that a Rayleigh instability could prevent the convergence 
of the Navier-Stokes solutions to Prandtl's solution was again suggested in \cite{CO10} for the case of the thick-core vortex. 
In \cite{OR90,CH02,CB06,KCH07}, the authors simulated the flow evolution induced
by a dipolar structure impinging on a no-slip boundary in which no adverse
pressure gradient is imposed at the beginning (as is the case in the present study). 
A shear instability developing small-scale structures was found to occur in the same Reynolds number 
range in which small-scale structures have been found to form in the present investigation ($Re\geq O(10^4)$).
The behavior of Navier-Stokes solutions at infinite Reynolds number is a longstanding 
problem in fluid dynamics. To determine whether the Navier-Stokes solutions converge to Prandtl solutions
close to the boundary and/or to Euler solutions away from the boundary is essential to advancing our understanding of high-Reynolds-number flows. 
It is impossible here to cite all the relevant results obtained in the
last several decades related to this very important issue.  
The reader interested in the mathematical theory of Prandtl's equations can see
the papers by \cite{CS00,WE00} and the book by \cite{OS99}.  See also the review paper by \cite{Cow01}. Regarding the
convergence of Navier-Stokes solutions in the zero viscosity limit,
we mention the papers by \cite{SC98a,SC98b,CS97,LCS01}  where, for analytic initial
conditions, the authors prove the convergence of Navier-Stokes solutions to
Euler and Prandtl solutions as $Re \to \infty$ for the flow in a half plane and in an exterior circular
domain. 
The result of \cite{SC98a} was later improved by requiring analyticity only in the streamwise direction in \cite{LCS03,CLS13,KV13}.
The convergence of the weak solutions of Navier-Stokes equations to Euler solutions has been considered 
in \cite{Kato84,TW97}, where the authors introduce criteria based on
\textit{a priori} estimates of energy dissipation and 
pressure gradient, respectively (see also \cite{CW07} for a discrete version of Kato's criterion, and \cite{Kel07} where an equivalent
condition based on the vorticity has been given for the convergence of Navier-Stokes to
Euler solutions). 
Strong convergence of Navier-Stokes to Euler solutions in $L^2$ spaces is given in
\cite{LML08} who impose symmetry properties. 
In \cite{CMR98,LLP05,Kel06,IP06},  
the authors study the inviscid limit of the two-dimensional Navier-Stokes equations in the case of a Navier friction 
boundary condition. 
In the next section, the physical problem, a two-dimensional circular cylinder
impulsively started in a uniform
steady background flow, is introduced, and the numerical schemes used to solve the Navier-Stokes equations are
presented (for the numerical scheme used to solve Prandtl's equation,
see \cite{GSS09}). In Section \ref{USP}, the various
viscous-inviscid interactions developing in both Prandtl and Navier-Stokes cases will be described.
In Section \ref{SA}, the methodology used to perform the
complex singularity analysis will be introduced. In particular, the singularity-tracking methods based on the
BPH method and the Pad\'{e} approximation will be explained. In Section
\ref{Wssa}, the investigation of the complex singularities for both Prandtl and Navier-Stokes
wall shear is carried out,
and the various stages of the unsteady separation process will be related to the presence of
different types of singularities.
In Section \ref{STMNS}, the complex singularity analysis is performed on the
streamwise velocity component of the Navier-Stokes solutions (the same study was performed in
\cite{GSS09} for Prandtl's solution).

\section{Statement of the problem and numerical schemes}
\label{state}
The case studied is the two-dimensional circular cylinder impulsively started in 
a uniform
steady background flow. We consider the reference frame fixed with the moving 
circular cylinder; therefore, the problem of the motion of the flow past a 
stationary
circular cylinder is considered. Cylindrical coordinates $(\theta,r)$ are used, 
where $\theta$ is the angular streamwise variable and $r$ is the normal variable 
to the circular cylinder.
Dimensionless variables are introduced taking the radius of the
circular cylinder $a$ and the uniform velocity $U$ of the flow at infinity
as characteristic length and velocity scales, respectively.
The relevant nondimensional parameter is the Reynolds number defined here as 
$Re=aU/\nu$, where
$\nu$ is the kinematic viscosity.
The governing equations for the flow evolution are the Navier-Stokes equations
in the domain $[0,2\pi]\times[1,\infty)$ of the form
\begin{eqnarray}
u_t+\frac{uu_{\theta}}{r}+vu_r+\frac{u v}{r}+\frac{p_{\theta}}{r}  =
\frac{1}{Re} \left( \frac{u_{\theta\theta}}{r^2}+\frac{u_{r}}{r}+u_{rr} 
-\frac{u}{r^2}+2\frac{v_\theta}{r^2}\right),
\label{NSequation_u}\\
v_t+\frac{uv_{\theta}}{r}+vv_r -\frac{u^2}{r} +p_r = 
\frac{1}{Re} \left( \frac{v_{\theta\theta}}{r^2}+\frac{v_{r}}{r}+v_{rr} 
-\frac{v}{r^2}-2\frac{u_\theta}{r^2}\right),
\label{NSequation_v}\\
\frac{u_{\theta}}{r}+\frac{v}{r}+v_r =  0. \label{NSequation_incomp}
\end{eqnarray}
Here, equations \eqref{NSequation_u} and \eqref{NSequation_v} are the
equations for the velocity components $(u,v)$,
\eqref{NSequation_incomp} is the incompressibility condition, and $p$ is the
pressure.
The boundary conditions are
\begin{eqnarray}
u(\theta,1,t)=v(\theta,1,t)=0, \label{noslipNS}\\
u(\theta,\infty,t)=\sin \theta , \quad v(\theta,\infty,t)=-\cos \theta,\\
u(0,r,t)=u(2\pi,r,t), \quad v(0,r,t)=v(2\pi,r,t).
\end{eqnarray}
The initial conditions for the velocity components are
\begin{equation}
u(\theta,r,0)=\psi^E_r \quad {\rm and} \quad
v(\theta,r,0)=-\frac{\psi^E_{\theta}}{r},
\end{equation}
where
\begin{equation}
\psi^E(\theta,r)= \left( r - \frac{1}{r} \right) \sin \theta \label{psi}
\end{equation}
is the streamfunction from the steady, inviscid Euler solution for this
configuration.

The no-slip boundary condition \eqref{noslipNS} imposed at the wall results in
vorticity generation on the circular cylinder, eventually leading to development of the unsteady separation phenomenon.
To describe the flow inside the boundary-layer, one defines the scaled
normal coordinate $Y$ and normal velocity $V$ by the well known boundary-layer
scaling $r=a + Re^{-1/2} \, Y$ and $v= Re^{-1/2} \, V$.
Prandtl's equations are obtained, to first order, by introducing the above
scaling into the Navier-Stokes equations and taking the limit as $Re \to \infty$.
For the impulsively-started circular cylinder, Prandtl's equations are
\begin{eqnarray}
\frac{\partial u}{\partial t}+u\frac{\partial u}{\partial x}+V
\frac{\partial u}{\partial Y}-U_{\infty}\frac{d
U_{\infty}}{d x} = \frac{\partial^2 u}{\partial Y^2},
\label{pramomentum}\\
\frac{\partial u}{\partial x}+\frac{\partial V}{\partial Y} = 0,\label{consmass}
\end{eqnarray}
with initial and boundary conditions given by
\begin{eqnarray}
u(x,Y,0)=U_{\infty},\label{initialprandtl}\\
 u(x,0,t)=V(x,0,t)=0, \quad
u(x,Y\rightarrow\infty,t)=U_{\infty},\label{boundaryprandtl}
\end{eqnarray}
where $U_\infty(x) = 2 \sin x$ is the inviscid Euler solution at the boundary.  The streamwise coordinate $x$ is measured along the cylinder surface from the front stagnation point, and the normal coordinate $y$ is measured from the cylinder surface.  Therefore, $(x,y) = (\pi - \theta, r - 1)$.

We solve the Navier-Stokes equations \eqref{NSequation_u}--\eqref{NSequation_incomp} in the
vorticity-streamfunction formulation, which is
\begin{eqnarray}
\frac{\partial \omega}{\partial t}+\frac{u}{r}\frac{\partial \omega}{\partial
\theta}+v\frac{\partial \omega}{\partial
r}=\frac{1}{Re} \left( \frac{1}{r^2}\frac{\partial^2 \omega}{\partial
\theta^2}+\frac{1}{r}\frac{\partial \omega}{\partial r}+\frac{\partial^2
\omega}{\partial r^2} \right), \label{NSequationd}\\
\frac{1}{r^2}\frac{\partial^2 \psi}{\partial \theta^2}+\frac{1}{r}\frac{\partial
\psi}{\partial r}+\frac{\partial^2 \psi}{\partial r^2}=-\omega,
\label{poissond}\\
u=\frac{\partial \psi}{\partial r},\quad v=-\frac{1}{r}\frac{\partial
\psi}{\partial \theta},\label{velocityewd}\\
u=v=0 , \qquad r=1, \label{noslippd}\\
\omega\rightarrow0, \qquad r\rightarrow\infty,\label{inftycond}\\
\omega(\theta,r,t=0)=0,\label{NSinitdd_vorticitybcd}\\
\psi(\theta,r,t=0)= \left( r - \frac{1}{r} \right) \sin \theta. \label{NSinitdd_PSI}
\end{eqnarray}

Equation \eqref{NSequationd} is the vorticity-transport equation, equation
\eqref{poissond} is the Poisson equation for the streamfunction, and
equations \eqref{velocityewd} relate the velocity components to the streamfunction. Boundary conditions
\eqref{noslippd} and \eqref{inftycond} are the no-slip and impermeability conditions on the circular cylinder and the irrotational
condition at infinity, respectively. The initial
condition \eqref{NSinitdd_vorticitybcd} expresses the irrotationality 
condition of the flow at the initial time, \eqref{NSinitdd_PSI} is the initial condition for the streamfunction.
The problem is solved in the domain $[0,\pi]\times[1,\infty)$, in which case only the upper
part of the circular cylinder is considered owing to symmetry, and then periodicity in the angular variable
is imposed.  

Given that the streamfunction $\psi$ becomes unbounded as $r\rightarrow\infty$, we
truncate the normal physical domain to a value $R_{max}$, where the vorticity
remains negligible for all computational times (let us say $\omega\leq10^{-16}$); therefore, the
computational domain is $D=[0,\pi]\times[1,R_{max}]$  . We adjust $R_{max}$
according to the Reynolds number accounting for the increasingly thin boundary layer with increasing Reynolds number.  Along with the other computational parameters, $R_{max}$ is reported in Table~1 for the various simulations.
To better resolve the boundary-layer region, where the more relevant phenomena
occurs, we have used a stretching function to cluster the computational grid
close to the boundary. This function is
\begin{equation}
 \Gamma (r)=\hat
r=1 + \frac{4}{\pi} \arctan \left[ b \tan \left( \frac{\pi}{2} \frac{r-R_{max}}{R_{max}-1.0} \right) \right],
\label{transformation}
\end{equation}
which maps the physical normal domain $1 \leq r \leq R_{max}$ onto the computational
domain $-1 \leq \hat{r} \leq 1$. The parameter $b>0$  determines  the degree of focusing of the
grid,
with a decreasing value of $b$ corresponding to an increased focusing close to
the boundary. Applying \eqref{transformation} to the Navier-Stokes
equations \eqref{NSequationd}--\eqref{inftycond}, we obtain the system of equations to be
solved in the domain $[0,\pi]\times[-1,1]$ as
\begin{eqnarray}
 \partial_t\omega=A(\hat{r})\partial_{\theta}\omega+B(\hat{r})\partial_{\hat{r}}
\omega+C(\hat{r})\partial_{\theta\theta}\omega+D(\hat{r})\partial_{\hat{r}\hat{r
}}\omega, \label{vorticityeq2}\\
E(\hat{r})\partial_{\theta\theta}\psi+F(\hat{r})\partial_{\hat{r}\hat{r}}
\psi+G(\hat{r})\partial_{\hat{r}}\psi=-\omega, \label{streameq}\\
\Gamma_{\hat{r}}(\hat{r})\partial_{\hat{r}}\psi=u,\quad
\Gamma(\hat{r}) \partial_{\theta}\psi=-v, \label{velocityeq2}\\
\omega(\theta,-1,t)=\xi(\theta,t), \quad \omega(\theta,1,t)=0,
\label{vorticitybound}\\
u(\theta,-1,t)=v(\theta,-1,t)=0, \label{ubound}\\
\omega(\theta,\hat{r},0)=0,\quad \psi(\theta,\hat{r},t=0)= \left( \Gamma^{-1}(\hat{r}) - \frac{1}{\Gamma^{-1}(\hat{r}) } \right) \sin \theta, \label{vorticityinit2}
\end{eqnarray}
where
\begin{eqnarray}
A(\hat{r})=-\Gamma(\hat{r}) u, \nonumber \\
B(\hat{r})=\frac{1}{Re}\Gamma_{\hat{r}\hat{r}}(\hat{r})-\Gamma_{\hat{r}}(\hat{r})v, \nonumber \\
C(\hat{r})=\frac{1}{Re} \Gamma^{2}(\hat{r}), \quad D(\hat{r})=\frac{1}{Re}
\Gamma_{\hat{r}}(\hat{r})^2, \nonumber \\
E(\hat{r})=\Gamma^{2}(\hat{r}), \quad F(\hat{r})=\Gamma_{\hat{r}}(\hat{r})^2, \nonumber \\
G(\hat{r})=\Gamma_{\hat{r}\hat{r}}(\hat{r}) + \Gamma(\hat{r}) \Gamma_{\hat{r}}(\hat{r}). \nonumber
\end{eqnarray}

To numerically solve the above system, a Galerkin-Fourier method is used 
in the angular variable, and the Chebyshev-Collocation method is used in the normal
variable.  This ensures fully spectral convergence (see \cite{Pey}). The temporal
discretization used is the  Adams-Bashforth-Implicit Backward Differentiation
(AD/BDI2) method, and to find the necessary vorticity boundary condition $\xi(\theta,t)$ at each time step, 
the Influence Matrix Method (\cite{Pey}) is used. Numerical solutions for
Reynolds numbers ranging from $10^3$ up to $10^5$ are computed, with computational
grids up to $8196\times1025$ points for the higher Reynolds numbers.
The parameter $b=0.1$ has been chosen in the transformation function (\ref{transformation}) for all computations. For all Reynolds numbers, the numerical
simulations are started using a coarse grid that is increasingly refined as the flow develops
small-scale structures.
Calculations for the higher Reynolds numbers were stopped when
solutions exhibited complicated behavior that required unachievable computational resolution.  
Recall the work of \cite{CW07} in which it was pointed out that the numerical solution of small-scale structures requires a
grid resolution that is of $O(Re^{-1})$. For example, spurious numerical oscillations have been observed in the present simulations 
for $Re=5\cdot10^4$ and $Re=10^5$ after formation of small-scale structures owing to the
lack of the necessary numerical resolution required to resolve these cases adequately. However, all of the
phenomena occurring during the separation process, which is the focus of our investigation,
are well resolved, and all the result are grid independent. 
To capture the asymptotic behavior of
the spectrum of the solution, in order to perform a reliable complex singularity analysis,  high numerical precision is required, and in fact in the
calculations we shall present, we have used 32-digit precision (\cite{BYLT02}). 

Finally, we refer the reader to \cite{GSS09} for all of the technical
details regarding the spectral numerical scheme used to solve, with high numerical precision, Prandtl's equations \eqref{pramomentum} and \eqref{consmass}.

\section{Unsteady separation process}
\label{USP}
\subsection{Prandtl's solution}\label{PS}
It has been known since Blasius's work (\cite{Bla08}) that Prandtl's solution develops a
singularity in a finite time in the case of the impulsively-started circular cylinder.
The physical mechanisms leading to the singularity formation are also well known (see for instance \cite{vDS80}).
In \cite{GSS09}, it has been shown that a cubic-root singularity arises as a shock forms in the
streamwise velocity component $u$. Therefore, our discussion of Prandtl's
solution will be brief and primarily focused
on those elements useful for comparison with Navier-Stokes solutions.

\begin{figure}
\includegraphics[width=13.5cm]{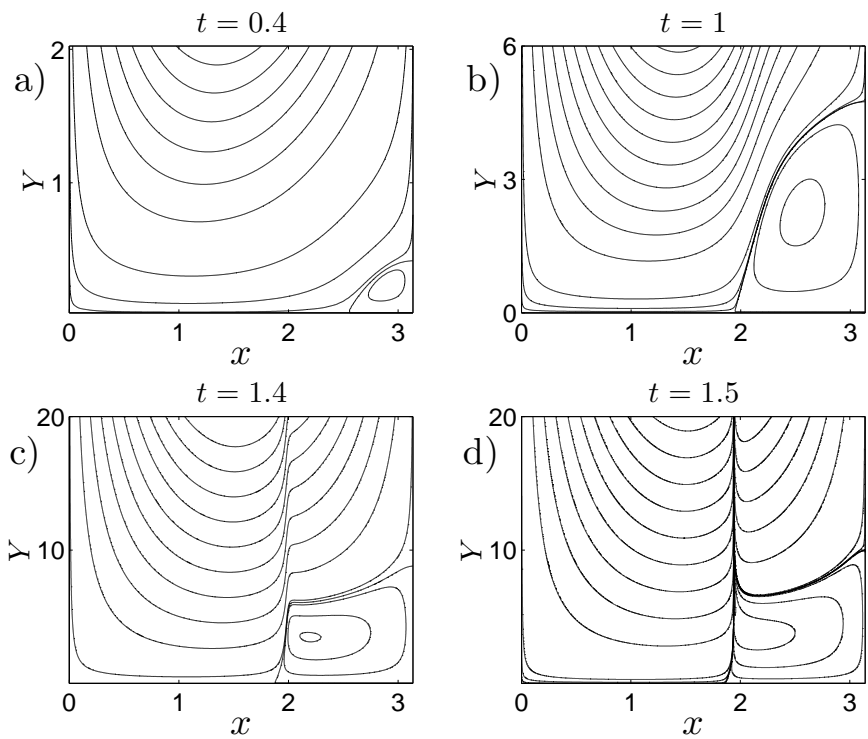}
\caption{Streamlines from Prandtl's solution at various times. A recirculation
region forms at $t_r\approx0.35$ and is visible at $t=0.4$. At $t_k=1.4$ a kink
forms above and to the left of the recirculation region, and rapidly evolves into a sharp spike at the singularity time $t_s=1.5$.}
\label{streampraDISK}
\end{figure}

\begin{figure}
\begin{center}
\includegraphics[width=10cm]{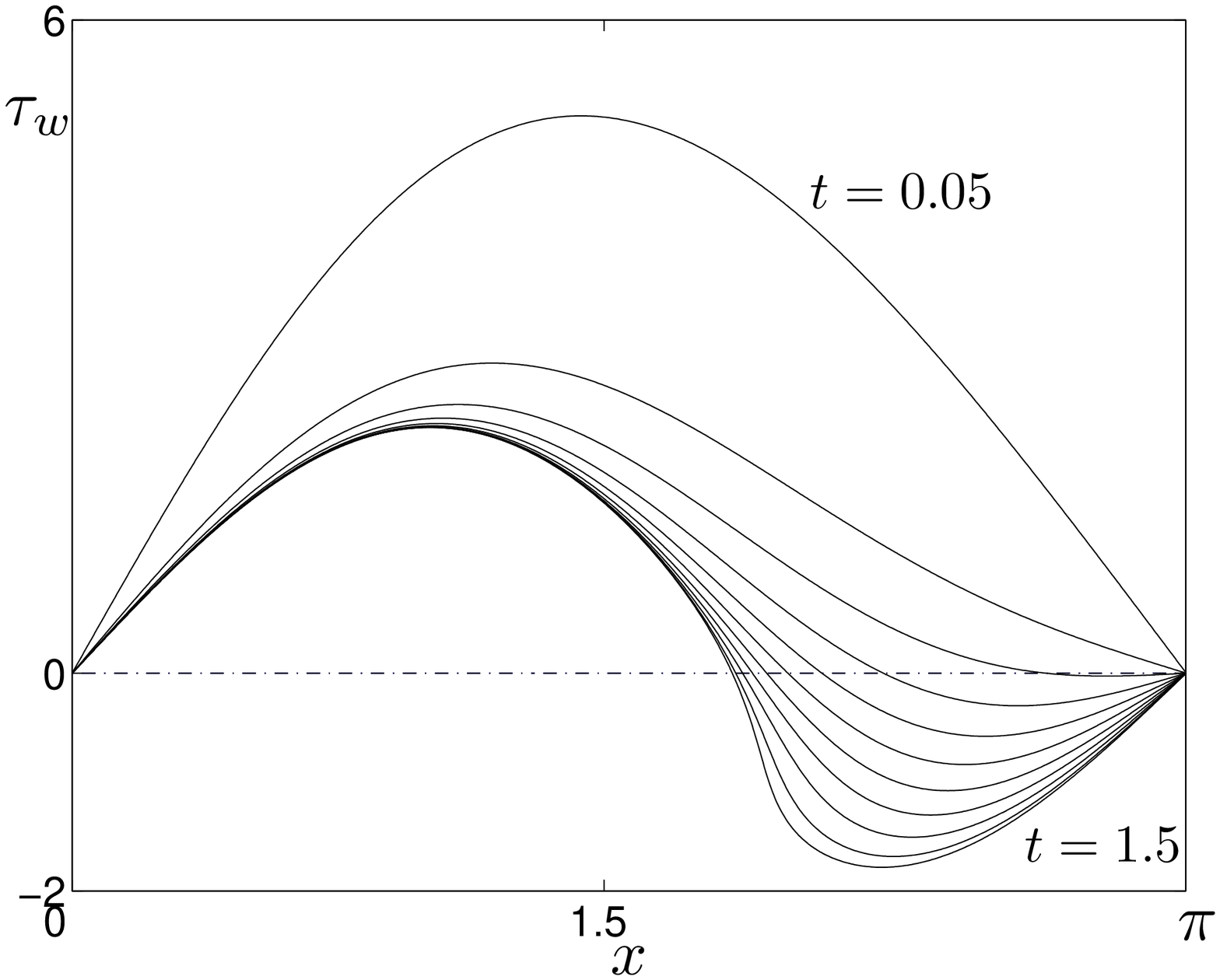}
\caption{Temporal evolution of Prandtl's wall shear $\tau_w^P$ from $t=0.05$ to
$t=1.4$ with temporal steps of $0.15$ and $t=1.5$.  At $t_r=0.35$, $\tau_w^P$
vanishes and a recirculation region forms. At $t_s=1.5$, the wall shear blows up in the second derivative at
$x_s\approx1.94$.}
\label{wall_shear}
\end{center}
\end{figure}

The primary factor leading to singularity formation is the presence of an adverse streamwise pressure gradient imposed
by the outer flow on a boundary layer. The adverse pressure gradient first leads to formation of a recirculation region attacked to the circular cylinder at $t_r\approx0.35$.
The formation of back-flow is clearly visible from the presence of closed
streamlines at time $t=0.4$ in Figure~\ref{streampraDISK}a.
The formation of the recirculation region also can be
inferred in this case from the vanishing of the wall shear $\tau_w^P=\partial_Y
u|_{Y=0}$ (see Figure~\ref{wall_shear}, where the temporal evolution of $\tau_w^P$ is shown).
In fact, for a flow that has positive wall shear everywhere and
only downstream motion at its initial time, the condition
$\tau_w^P(x,t)=0$ signals the onset of reversed flow within the boundary layer.

The first point of zero wall shear moves rapidly upstream,
which defines the upstream location of the growing recirculation region on the
downstream side of the circular cylinder.
At approximatively $t_k\approx1.4$, a kink forms in the streamlines and vorticity
contour levels owing to the pressure gradient that forces the fluid to deflect
upward away from the surface as shown in Figure~\ref{streampraDISK}c.  
According to the interpretation given in \cite{PSW91a,PSW91b}, the
formation of the kink represents the first stage of the viscous-inviscid interaction in the boundary
layer.
In fact, for $t<t_k$ the normal thickness of the boundary layer is of
the same order
as the boundary-layer scale.  Physically, therefore, the boundary layer remains
thin on the circular cylinder. For $t>t_k$, the fluid particles are pushed
away from the
boundary, and the boundary layer rapidly focuses
in a very narrow zone on the left of the recirculation region. At
$t_s\approx1.5$,
the kink in the streamlines and vorticity contour levels becomes a sharp spike near
$x_s\approx1.94$, revealing the singularity formation in the solution
and the consequent breakdown of the boundary-layer assumptions.
In \cite{GSS09}, the singularity formation for Prandtl's equation has been studied
through
the singularity-tracking method (see Section \ref{sing_methods}),
and it has been shown that for the initial condition $U_{\infty}(x)=\sin x$,
a cubic-root singularity forms at $t_s=3$
with the blow up of $\partial_x u$ at $Y\approx7$ (note that for the initial
condition
given by \eqref{initialprandtl}, the singularity forms at $t_s=1.5$ at
$Y\approx5$).

\begin{figure}
 
\begin{center}
\includegraphics[width=8cm]{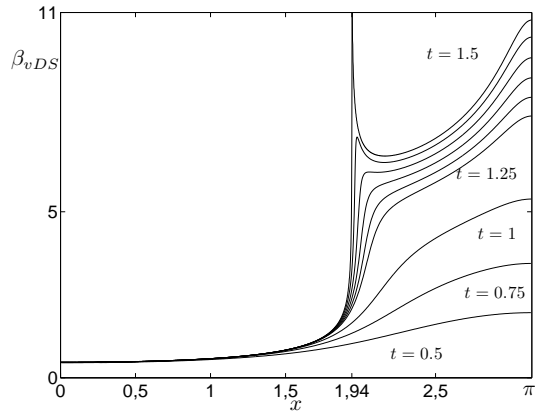}
\caption{Temporal evolution of the displacement thickness $\beta_{vDS}$ from
$t=0.5$ to $t=1.0$ with temporal steps of 0.25, and from
$t=1.25$ to $t=1.5$ with temporal steps of 0.05. At $t_k\approx1.4$, the same time
at which the kink in the streamlines forms
(see Figure~\ref{streampraDISK}c), a local maximum forms, and this leads to the
viscous-inviscid interaction that is followed by the break-up
of the boundary-layer assumption and the blow-up of $\beta_{vDS}$ at $t_s=1.5$.}
\label{displacement}
 
\end{center}
\end{figure}

The growth of the boundary layer can also be illustrated through the
displacement thickness, which is defined by
\begin{equation}
 \beta_{vDS}=\int_0^{\infty}1- \frac{u(x,Y,t)}{U_{\infty}(x)}dY.
\end{equation}
The temporal evolution of the displacement thickness is shown in
Figure~\ref{displacement}. 
At $t\approx t_k$, a local maximum forms in the displacement thickness,
and this signals the onset of the interaction between the viscous boundary-layer flow
and the inviscid outer flow. The displacement thickness abruptly focuses in a narrow
zone close to
$x_s\approx1.94$, and at $t=t_s$ it blows up revealing the singularity
formation.

%


\subsection{Navier-Stokes solutions}\label{NSS}

In this section, we shall describe the behavior of solutions of the
Navier-Stokes equations at different
Reynolds numbers ($10^{3} \leq Re \leq 10^{5}$). Specifically, comparisons between
Navier-Stokes solutions and Prandtl's solution will be provided that primarily focus on the characterization of the large- and small-scale interactions occurring during the separation process as carried out in \cite{Cas00,OC02}
for the thick-core vortex and later in \cite{GSS11} for the
rectilinear vortex. The large-scale interaction,
which is manifest for all finite Reynolds numbers, represents the first reaction of the
inviscid outer flow to the formation of the viscous boundary layer. During this
stage, the first relevant discrepancies
between Navier-Stokes and Prandtl's solution arise.
The small-scale interaction, on the other hand, is only manifest for moderate to high Reynolds 
numbers, and it coincides with formation of large
streamwise gradients, formation of various small-scale structures in
the flow, and kink formation in the streamlines and vorticity contours.
We now briefly explain the main events characterizing these interactions for the
various Reynolds numbers considered.  
Refer to \cite{Cas00,OC02,GSS11} for a more exhaustive treatment
of this topic.

\begin{figure}
\begin{center}
\subfigure[$Re=10^3$]{\includegraphics[width=8cm]{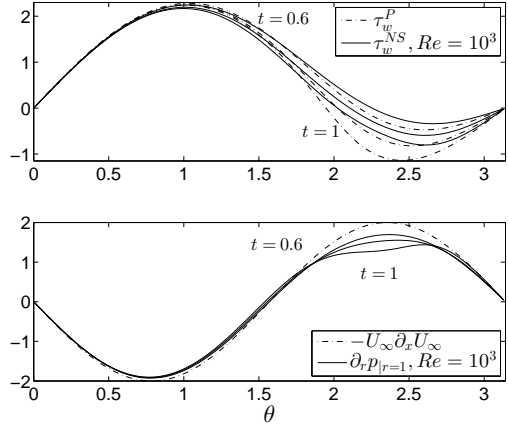}}
\subfigure[$Re=10^5$]{\includegraphics[width=8cm]{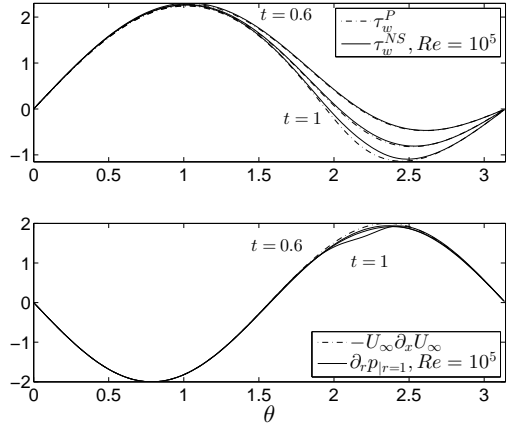}}
\caption{A comparison between the wall shears $\tau_w^P$ (dotted) and $\tau_w^{NS}$ (dashed) 
for a) $Re=10^3$ and b) $Re=10^5$ at time $t=0.6,0.8,1$.
A comparison between streamwise pressure gradient
$\partial_{\theta}p_{|r=1}$ (dashed) and streamwise pressure gradient of
Prandtl's solution (dotted) for  a) $Re=10^3$ and b) $Re=10^5$ at time $t=0.6,0.8,1$\ldots. At time
$t=1$, the differences between
Prandtl and Navier-Stokes solutions are clearly visible for both $Re=10^3$ and $Re=10^5$ owing to the
large-scale interaction.}
\label{comparison_t_p}
\end{center}
\end{figure}

\begin{figure}
 
\begin{center}
 \includegraphics[width=8.5cm]{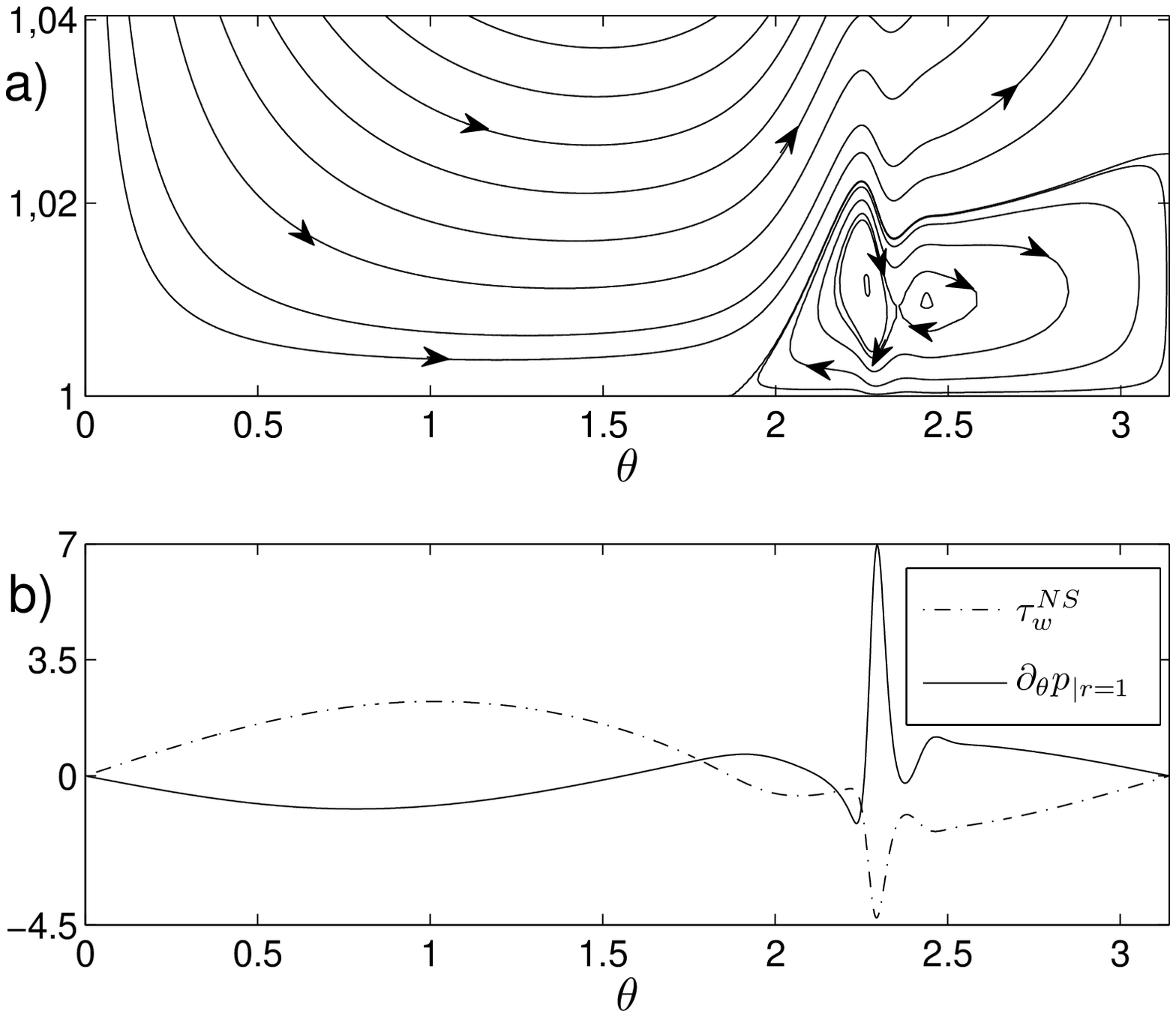}
\caption{a) The streamlines for $Re=10^5$ at $t=1.4$. b)
The wall shear (dotted) and angular pressure
gradient on the circular cylinder (dashed and rescaled by a factor 2) at $t=1.4$. At this
time, small-scale interaction causes 
formation of large gradients in $\tau_w^{NS}$ and $\partial_{\theta}p_{|r=1}$, a kink in
the streamlines above and to the left of the primary
recirculation region, and splitting of the recirculation region.}
\label{streamDISK105p4}
\end{center}
\end{figure}

\begin{figure}
\subfigure[$Re=10^3, t=1.1$]{\includegraphics[width=7.0cm]{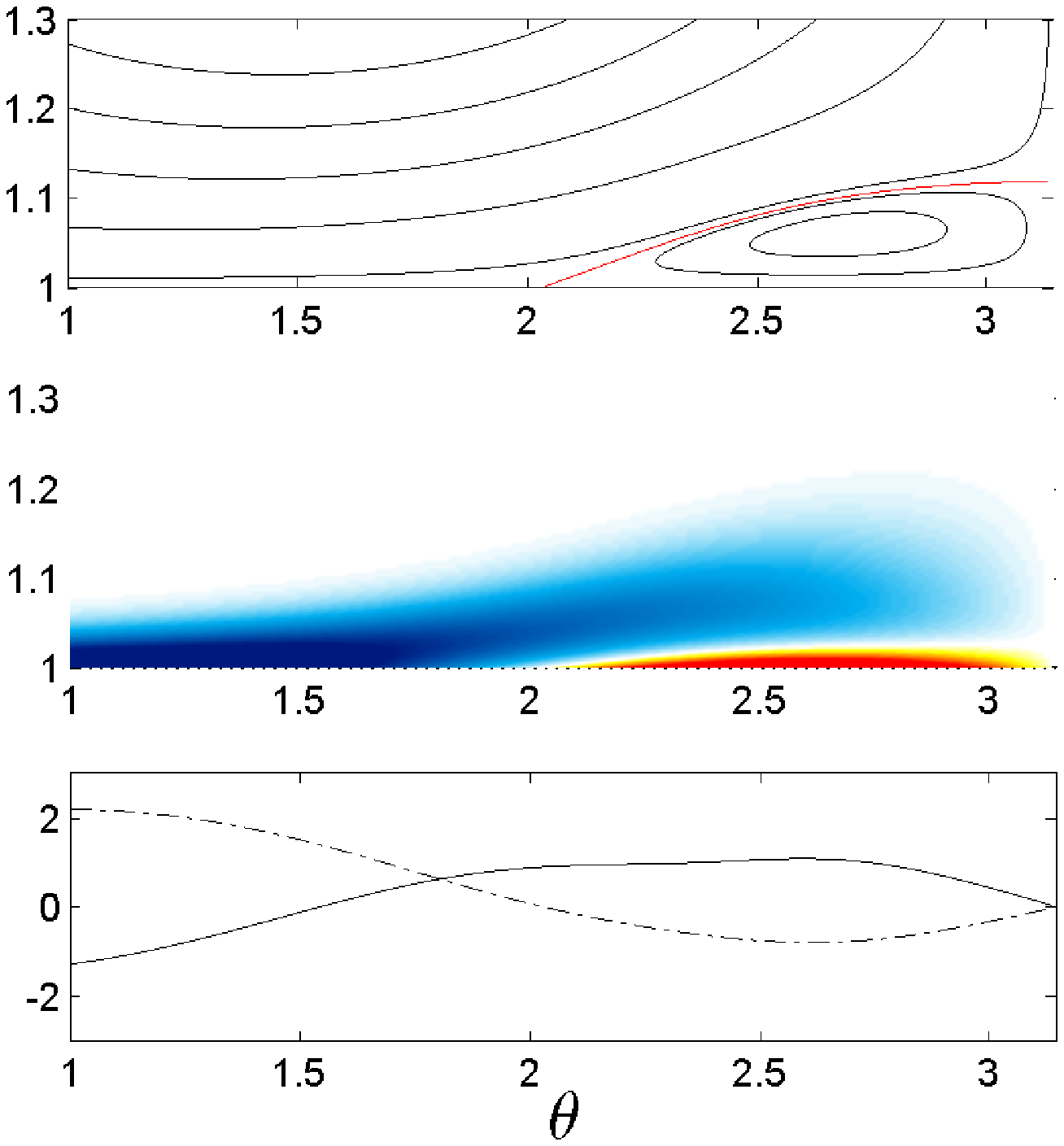}}
\subfigure[$Re=10^5, t=1.1$]{\includegraphics[width=6.9cm]{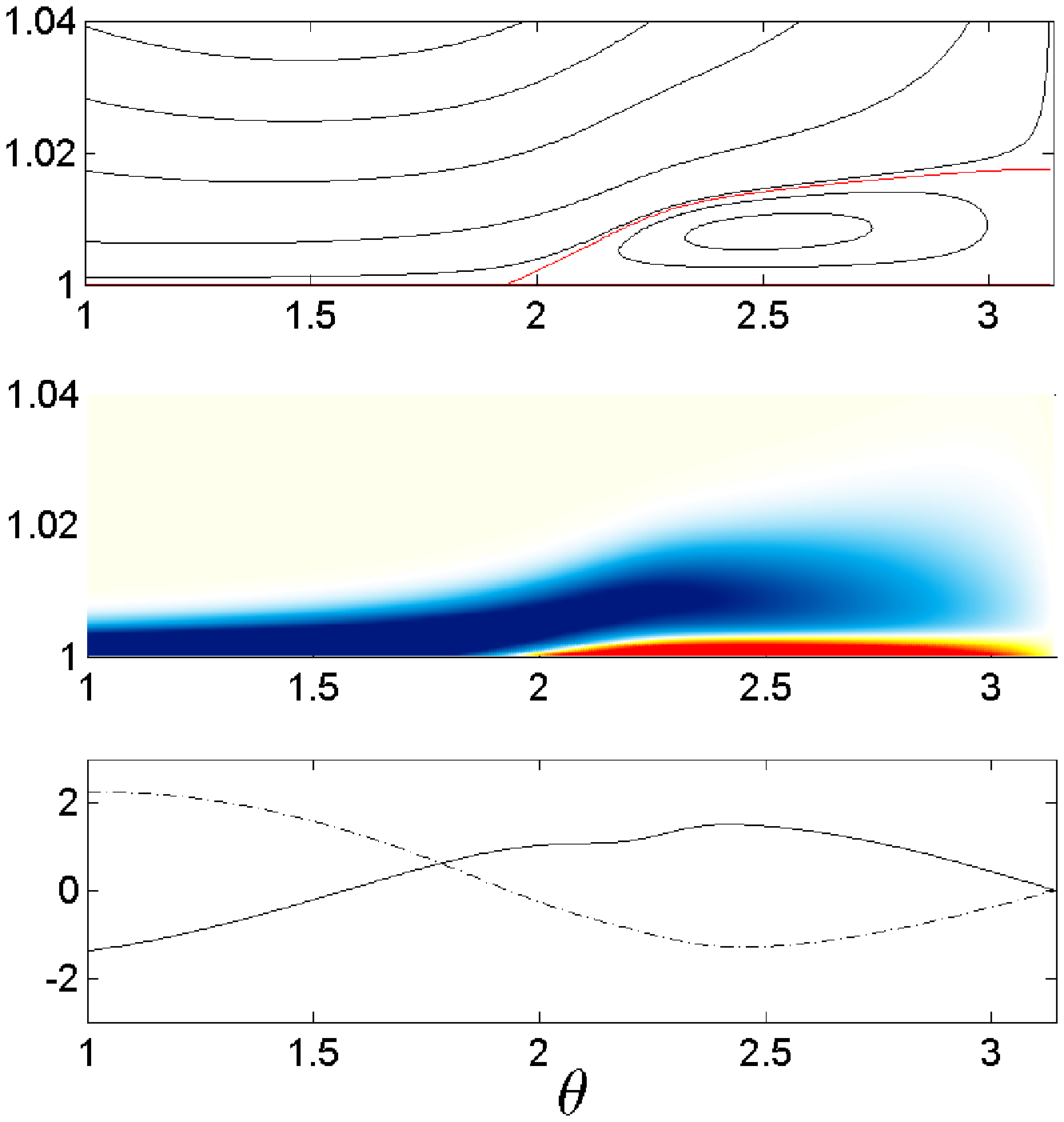}}
\caption{Upper figures: the streamlines (zero levels in red). Middle figures:
vorticity contour levels (blue colors negative vorticity,
red colors positive vorticity). Lower figures: $\partial_{\theta}p_w$ (dashed)
and  $\tau_w$ (dotted).}
\label{10p310p5ls}
\end{figure}

During the first stage of the separation process, i.e.\ before the formation of the
large- and small-scale interactions,
the flow evolution is qualitatively similar for all Reynolds numbers and 
agrees with that predicted by Prandtl's solution corresponding to $Re \to \infty$.
The most relevant physical event characterizing this stage is the
formation of the recirculation region,
which forms at time $t_r\approx0.35$ as in Prandtl's case. Moreover, comparison
of the wall shear stress and the streamwise pressure gradient shows very good agreement as one can see in Figure~\ref{comparison_t_p} for $Re=10^3$ and $Re=10^5$ at $t=0.6$.
The first noticable differences between the Navier-Stokes and Prandtl solutions
can be detected by the local change of $\partial_{\theta}p_w$ and $\tau_w^{NS}$ in the Navier-Stokes solutions as compared to  the same quantities from Prandtl's solution.
These changes are quite evident after time $t\approx0.9$ for all Reynolds numbers
considered, as one can see in Figure~\ref{comparison_t_p}
at time $t=1$ for $Re=10^3$ and $Re=10^5$; this corresponds to the beginning of the
large-scale interaction. We shall define the beginning of large-scale
interaction in the same way as in \cite{GSS11} for the
rectilinear vortex, according to which the
large-scale interaction begins when an inflection point forms on the left of the
maximum of $\partial_{\theta}p_w$.  In Figure~\ref{comparison_t_p},
the change of concavity close to the maximum in the streamwise pressure gradient
is visible at $t=1$ for both $Re=10^3$ and $Re=10^5$.
This inflection point carries a physical meaning, as it is the precursor to the
formation of a local minimum in the pressure gradient
that eventually becomes negative and therefore reflects the formation of an
adverse pressure gradient under the primary recirculation region.  This 
leads to formation of a secondary recirculation region attached to the
circular cylinder. In Figure~\ref{streamDISK105p4},
this local negative minimum in $\partial_{\theta}p_w$ is visible close to
$\theta\approx2.25$ at $t=1.4$ for $Re=10^5$.
The times $T_{LS}$ of formation of this inflection point and the angular
locations $\theta_{LS}$ where it forms
are reported in Table~2 for all cases. 
\begin{table}
\begin{center}
    \begin{tabular}{lllllllll}
    \hline
    $Re$ & $T_{LS}$ & $\theta_{LS}$ & $t_{p}$ & $\theta_{p}$ & $t_{w}$ &
$\theta_{w}$ & $t_{ss}$ & $\theta_{ss}$ \\
\hline
    $10^3$ & $0.908$ & $2.308$  \\
$10^4$ & $0.916$ & 2.25 & $1.512$ & 2.65 & 1.55 & 2.68 & 1.505 & 2.62\\
$5\cdot10^4$ & $0.94$ & 2.21 & $1.31$ & 2.48 & 1.355 & 2.49 & 1.29 & 2.47\\
$10^5$ & $0.952$ & 2.18 & $1.3$ & 2.45 &1.315 & 2.45 &1.26 & 2.419\\
\hline
 \multicolumn{5}{c}
	{\rule[-3mm]{0mm}{8mm}}
  \end{tabular}
\end{center}
\caption{The time and location at which large-scale interactions begins,
i.e the inflection point forms in $\partial_{\theta}p_{r=1}$
($T_{LS},\theta_{LS}$),
the local minimum in $\partial_{\theta}p_{r=1}$ forms at ($t_{p},\theta_{p}$),
the local maximum in $\tau_w^{NS}$ forms at ($t_{w},\theta_{w}$),
small-scale interaction begins, i.e the real location of the complex singularity
$s_{ss}$
begins to move upstream on the circular cylinder, at ($t_{ss},\theta_{ss}$).}
\label{largescaletable}
\end{table}
We shall see through the singularity analysis
performed in
Sections~\ref{Wssa} and \ref{STMNS} that at time $T_{LS}$, some relevant changes
in the complex singularities of $\tau_w^{NS}$ and $u$
can be detected.
It is evident that the formation of large-scale interaction occurs
earlier as Reynolds number decreases, and the location of the inflection point moves
upstream on the circular cylinder
as Reynolds number increases.
However, even if some discrepancies are observed between Prandtl and
Navier-Stokes solutions, the qualitative flow behavior is similar for all Reynolds numbers considered, and they agree with that prescribed
by Prandtl's solution.  In fact,
only one recirculation region is present as one can see in 
Figure~\ref{10p310p5ls}a,b, where the streamlines are shown for $Re=10^3$ and $Re=10^5$
at $t=1.1$.

After large-scale interaction begins, the flow evolution is strongly dependent on the
Reynolds number,
and two different regimes can be identified: moderate to high Reynolds numbers ($Re\geq
O(10^4)$), for which
the unsteady separation process is characterized by the small-scale interaction,
and a low-Reynolds-number regime,
for which small-scale interaction does not evolve after the large-scale
interaction.
To understand the effect of this new interaction, consider Figure~\ref{streamDISK105p4},
where the streamlines are shown and compared with $\partial_{\theta}p^w$ and
$\tau_w^{NS}$ at time $t=1.4$ for $Re = 10^5$.  In this figure, a kink located above and to the left of the recirculation
region is clearly visible as the result of the strong compression in the
near-boundary region.  This compression also leads to 
splitting of the recirculation region, and a second small recirculation region
is visible on the right of the primary recirculation region.
In correspondence to the split recirculation region, one can also observe the
strong streamwise variations in $\partial_{\theta}p_w$ and $\tau_w^{NS}$. The
kink
rapidly evolves into a spike, and it is responsible for the growth of the boundary layer in the normal
direction. This
physical behavior resembles the singularity formation in Prandtl's solution, and
this is the physical characterization of the small-scale interaction that is
visible
only for moderate to high Reynolds numbers ($Re\geq10^4$ in our case). We observe 
that, as seen in \cite{OC02,GSS11}, the large-scale interaction rapidly
evolves into a 
small-scale interaction as Reynolds number increases.
A plausible start time for the onset of the 
small-scale interaction will be given 
by considering the physical phenomena characterizing this interaction in terms of
the complex singularity analysis of $\tau_w^{NS}$ and $u$.

For low Reynolds number, e.g.\ $Re=10^3$, no small-scale interaction
develops. In fact, there is no evidence of any kink formation in the streamlines,
splitting of the primary recirculation region,
or formation of large gradients. We have numerically simulated the case with $Re=10^3$ up to $t=6$, well after detachment of the boundary layer,
and no evidence of small-scale interaction is
detected. The different flow evolutions observed for $Re=10^3$ can be explained
owing to the more pronounced diffusive effects acting for low Reynolds number that prevents the
strong compression leading to kink and spike formation.

In \cite{GSS11}, it also has been shown how the small-scale interaction
strongly influences the enstrophy evolution $\Omega=\lVert\omega \rVert_{L^2(D)}^{2}$,
where $D$ is the boundary layer region
($D$ is chosen so that the vorticity outside $D$ remain negligible
for all computational time). The enstrophy represents the energy decay rate
according to the temporal laws
\begin{eqnarray}
\frac{dE(t)}{dt}&=&
-\frac{1}{Re}\Omega(t)+ NT_1, \label{energy}\\
\frac{d\Omega(t)}{dt}&=&-\frac{2}{Re}P(t)+
2I^p(t)+NT_2, \label{enstrophy}
\end{eqnarray}
where $E=\frac{1}{2}\lVert \textbf{u} \rVert_{L^2(D)}^{2}$ and $P=\lVert\nabla^{r,\theta}\omega
\rVert_{L^2(D)}^{2}$ are the energy and palinstrophy
within the boundary layer $D$,
and
\begin{eqnarray}
I^p(t)&=& \int_{0}^{2\pi}\omega_{|r=1}\cdot\partial_{\theta} p_w
d\theta, \nonumber
\end{eqnarray}
where $\mathbf{n}$ is the exterior normal to $\partial D$ and 
the $NT_i$ are negligible terms.
%

\begin{figure}
 
\begin{center}
\includegraphics[width=10.5cm]{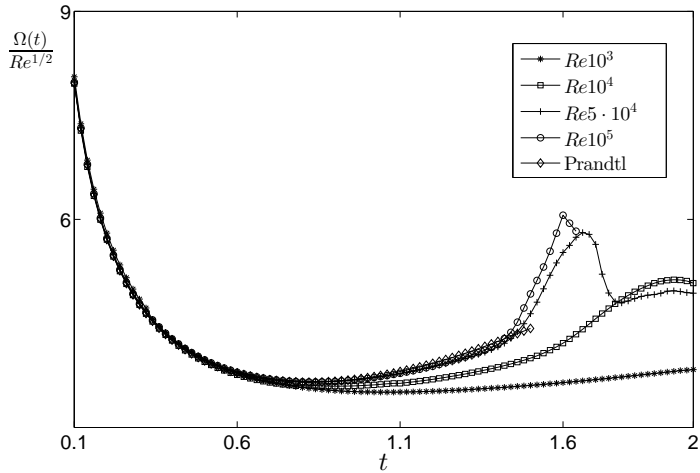}
\caption{The temporal evolution of the rescaled Navier-Stokes enstrophy
$\Omega=\lVert\omega \rVert_{L^2(D)}^{2}Re^{-1/2}$
and Prandtl's enstrophy $\lVert\partial_Y u \rVert_{L^2}^{2}$. Prior to the onset of large-scale
interaction, the Navier-Stokes enstrophy agrees well with Prandtl's enstrophy. During the small-scale interaction, the
small-scale vortical structures forming within the boundary layer increase the enstrophy owing to the large
amount of vorticity produced on the cylinder surface.}
\label{enstrofia}
 
\end{center}
\end{figure}

During the small-scale interaction stage, several dipolar structures
form in the boundary layer, and during their impingement on the circular cylinder, a large
amount of vorticity is produced, leading to growth of enstrophy. In
Figure~\ref{enstrofia}, the temporal evolution of the rescaled enstrophy $\Omega Re^{-1/2}$ of the Navier-Stokes solution is shown and compared to the enstrophy $\lVert\partial_Y u \rVert_{L^2}^{2}$ computed
from Prandtl's solution. 
Up to the time at which large-scale interaction begins, the enstrophy of the Navier-Stokes
solutions agrees closely with that from Prandtl's solution for all the Reynolds numbers considered. For moderate to high Reynolds
numbers, the enstrophy grows owing to formation of
small-scale vortical structures within the boundary layer that begin to
interact between them and to impinge on the circular cylinder
creating a large amount of vorticity production.
For $Re=10^3$, it is found that the only effect that increases the
enstrophy is that owing to the primary recirculation, which after its total detachment
from the circular cylinder, gets close once again to the circular cylinder. Similar behavior has been shown previously
in \cite{CB06,KCH07}, in which the authors numerically simulate the
interaction of a vortex dipole
with a no-slip boundary. They found a similar range of Reynolds numbers
($Re\geq O(10^4)$) for which small-scale interaction develops within the
boundary
layer, and the enstrophy evolution shows peaks during the impingement of the various dipolar structures on the wall.
Because the flow evolution observed for the impulsively-started circular cylinder has
many similarities
to that simulated in \cite{CB06,KCH07,GSS11}, we refer the reader
to these papers for a more
exhaustive discussion on the influence of the small-scale interaction on the temporal evolution of the
enstrophy.


\section{Singularity analysis}
\label{SA}

The phenomena characterizing the unsteady unsteady separation
process in both Prandtl and Navier-Stokes cases are evaluated in this section by performing an analysis of complex singularities
in their solutions.
In particular, we focus on wall shear stress in both cases.
Because the wall shear acts to increase the rate of enstrophy production (see the previous
definition of $I^p$) and it is 
a strong indicator of the various regimes forming in the separation process, it
is of interest 
to perform the analysis of its singularities to check if they can be related to the various stages and regimes within
the separation process.
In the Navier-Stokes case, it is also natural to perform a similar analysis on the streamwise
pressure gradient along the surface of the circular cylinder. Recall that
in Prandtl's case, however, the streamwise pressure 
gradient is analytic and imposed by the outer flow; therefore, no complex singularities
are present.  Therefore, our focus will be on analysis of the wall shear stress.

Before showing the results of this
analysis, the methods used to perform the singularity tracking will be described.
Moreover, a bi-dimensional analysis will be performed on the velocity component
$u(r,\theta)$ of the Navier-Stokes solution similar to that performed in \cite{GSS09} for Prandtl's solution, and we shall see how the complex singularities
can be related to the various stage of unsteady separation discussed in the previous
section.

\subsection{Singularity analysis: methods}
\label{sing_methods}

To analyze the complex singularities of the wall shear, most of
the methods currently used for such studies of
one-dimensional functions $u(z)$ expressed as a Taylor or
Fourier series have been considered. Brief explanations are given for each of these methods, and the reader is referred to the extensive literature that will be cited for a deeper
understanding on the theory behind these methods.

The first method used is generally referred to as a singularity-tracking method,
and it allows one to characterize the singularity of $u(z)$ through
analysis of its Fourier spectrum. In particular, given 
$u(z)=\sum_{k=-K/2}^{k=K/2}u_k\textrm{e}^{ikz}$ with a complex singularity
at
$z^* = x^* + i\delta$ and $u(z) \approx (z-z^*)^\alpha$ as $z\rightarrow z^*$, 
the asymptotic behavior of its spectrum is governed by Laplace's
formula (see \cite{CKP66})
\begin{equation}
u_k\sim
|k|^{-(1+\alpha)}\exp{(- \delta |k|)}\exp{(ix^*k)} \quad \mbox{as} \quad k \rightarrow \infty.
\label{asyspectrum}
\end{equation}
If one is able to estimate the rate
of exponential decay $\delta$ of the spectrum, the
distance of the complex singularity from the real axis can be obtained.
The estimate of the period of the oscillations of the spectrum gives
the real location $x^*$ of the singularity.
Resolving the rate of
algebraic decay $1+\alpha$, one can then classify the singularity type.
This method has been used extensively to track the complex singularities for both
ordinary and partial differential equations (see \cite{SSF83,Sh92,Caf93,GPS98,CBT99,FMB03,MBF05,PMFB06,DLSS06,PF07,GSS09,CGS12}.  
The primary drawback of this method is the fact that
it gives no information about complex singularities located outside the
width $\delta$ of the analyticity strip. This method is generally used along with
robust fitting procedures like sliding fitting (see \cite{Sh92,Caf93,DLSS06,GSS09}) .  
This approach requires high numerical precision in the simulation to
avoid interference of the round-off error that is usually present when one deals with
Fourier spectra.

\begin{figure}
 
\begin{center}
\includegraphics[width=6.0cm]{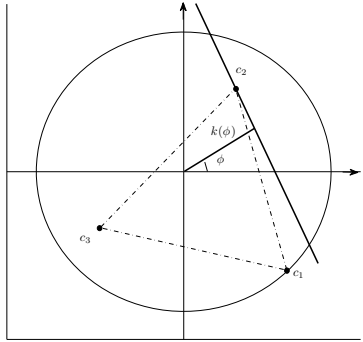}
\caption{The convex hull of a discrete set of complex singularities is the
smallest convex polygon containing all the singularities $c_i$.
The \textit{supporting} function $k(\phi)$ is the distance from the origin to
the supporting line, normal to the direction $\phi$, and 
touching a singularity.}
\label{indicatrix}
 
\end{center}
\end{figure}

To retrieve more information about the possible singularities outside the width
of the 
analyticity strip, the BPH (Borel-P\'{o}lya-Hoeven) method proposed in
\cite{PF07} can be used.  
The authors perform an analysis on the complex singularities 
of Burgers equation for different initial conditions through the asymptotic behavior of the 
Borel transform of a Taylor series. This method can be used when one deals with a finite
number of distinct complex singularities (poles or branches) as actually happens
in wall shear in both Prandtl and Navier-Stokes contexts.
In particular, given the inverse Taylor series $u(z) = \sum_{k=0}^N u_k/z^{k+1}$
that has $n$ complex singularities $c_j=
|c_j|$e$^{-i\gamma_j}$ for 
$j=1,2,\ldots, n$, its Borel transform is given by $U_B(\zeta)=\sum_{k=0}^N
u_k\zeta^{k}/n!$.
Evaluating the modulus of the Borel series $G(r)=|U_B(r\textrm{e}^{i\phi})|$
along the rays $r\textrm{e}^{i\phi}$, one obtains, through a steepest descent
argument, the following asymptotic behavior
\begin{equation}
G(r)\approx C(\phi)r^{-(\alpha(\phi)+1)}\textrm{e}^{h(\phi)r} \quad
\textrm{for}\quad r \rightarrow \infty.
\label{borel_as}
\end{equation}
Here, the function $h(\phi)$ is called the \textit{indicatrix} function of the Borel
transform. To better understand the role
of the indicatrix function, the set $K=\{c_1,\ldots,c_n\}$ of
all the singularities is considered, and the \textit{supporting line} of $K$ is defined
by a line that has at least one point in common with $K$ and such that its points
are in the same half space with respect to the supporting line of $K$.
The intersection of all these half spaces is the \textit{convex hull} of $K$,
which in the case of separate poles or branches reduces to the smallest
convex polygon containing all the singularities as illustrated in Figure~\ref{indicatrix}.
The \textit{supporting} function $k(\phi)=h(-\phi)$ is the distance from the
origin to the supporting line normal to $\phi$.
In \cite{PF07}, it has been shown how, in the case of isolated singularities, the
function $h$ varies along the angular direction $\phi$ as 
\begin{equation}
h(\phi)=|c_j|\cos(\phi-\gamma_j) \quad  \textrm{for} \quad
\phi_{j-1}<\phi<\phi_j,
\end{equation}
where the set of angular directions $\phi_j$, $j=1,2,\ldots,n$ is determined by the
angle $\phi$,
for which the supporting line normal to $\phi$ touches $K$ in $c_j$ (see
Figure~\ref{indicatrix}).
Therefore, the indicatrix function $h(\phi)$ is a piecewise cosine function, and
through numerical interpolation we can determine
the parameters $|c_j|$ and $\gamma_j$ that give the locations of the complex
singularities $c_j$. In practice, for each direction $\phi$ we need to determine
the exponential rate of \eqref{borel_as} that allows for construction of the indicatrix
function $h$. Moreover, an estimate of $\alpha(\gamma_j)$ in \eqref{borel_as} 
returns the characterization of the singularity $c_j$. The BPH method easily can be applied to the Fourier series $u(z)=\sum_{k=-K/2}^{k=K/2}u_k\textrm{e}^{ikz}$
by writing $u$ as a Taylor series. 
The advantage of this methodology in comparison to the singularity-tracking
method lies in the fact that
it is possible to capture information on all the singularities located in the
convex hull
outside the radius
of convergence of a Taylor series (or the strip of analyticity of a
Fourier series). However, there are some
drawbacks. In particular, singularities that are close to each other can be
difficult to distinguish if only a few terms are used in the
Borel series. Moreover, the computational cost
is heavier in comparison to the singularity-tracking method, as a numerical
interpolation must be performed in various directions containing all of the singularities.

The third method used is based on Pad\'{e} approximations. Suppose there is a
complex
function $u(z)$ expressed by a power series $u(z)=\sum_{k=0}^{\infty}u_iz^k$.
The Pad\'{e} approximant $P_{L/M}$
is a rational function approximating $f$, such that
\begin{equation}
f(z)\approx\frac{\sum_{i=0}^{L}a_i z^i}{1+\sum_{j=1}^{M}b_j z^j}=P_{L/M},
\label{pade}
\end{equation}
where $L+1$ and $M$ are the number of coefficients in the numerator and
denominator, respectively. The $M$ unknown denominator coefficients $b_j, j=1\ldots,M$, are first 
determined uniquely by equating coefficients of equal powers of $z$ between
$(\sum_{i=0}^{\infty}c_iz^i) (1+\sum_{j=1}^{M}b_j z^j)$  and $\sum_{i=0}^{L}a_i
z^i$, 
setting the coefficients of order greater than $L$ equal to zero, and $b_0=1$
by definition.
The following  set of $M$ linear equations must then be solved
\begin{equation}
\begin{array}{c}
b_Mc_{L-M+1}+b_{M-1}+\ldots+b_0c_{L+1}=0,\\
\vdots \\
b_Mc_L+b_{M-1}c_{L+1}+\ldots+b_0c_{L+M}=0.
\end{array}
\label{padematrix}
\end{equation}
Then the $L+1$ unknown numerator coefficients $a_i, i=0,\ldots,L$ 
follow from $(\sum_{i=0}^{\infty}c_iz^i) (1+\sum_{j=1}^{M}b_j
z^j)=\sum_{i=0}^{L}a_i z^i$ by equating coefficients of equal powers of $z$
less then or equal to $L$.

The advantage of the Pad\'{e} approximation method is that it allows one to continue the
function $f$ even beyond the radius of convergence of the Taylor series
$\sum_{i=0}^{\infty}c_iz^i$, and one only has the difficulty of convergence 
near branch points or branch cuts of $f$.
Thus, the approximation $P_{L/M}$ is able to represent all the singularities of
$f$ by
detecting the zeros of the denominator of $P_{L/M}$. 
The disadvantage of the Pad\'{e} approximation method is that not all of the singularities
represented by a general $P_{L/M}$ are singularities of the function being
approximated.
In fact, there are several examples (see, for example, \cite{BGM96}) for which some defects or
spurious singularities can appear. 
However, these defects can in principle be detected as they generally manifest themselves as a
pole very near to zeros in $P_{L/M}$.  Fortunately, these unusual occurrences have a transient nature that can be
neglected as they generally appear or disappear by changing the degrees of the
Pad\'{e} approximation. Note that the linear system (\ref{padematrix}) is close to being
singular, particularly for a high degree of the
approximant, and using high numerical precision
can in part overcome this issue. The results presented here have been tested to be
free from such spurious results or defects. Pad\'{e} approximants also have been used in the
analysis of complex singularities
of various ordinary differential equations (see Weideman 2003). The theoretical and practical issues related to
Pad\'{e}-based methods are so numerous that it is
impossible to cite them all here, and the reader is referred to \cite{BGM96} for a good
discussion of this topic. 
Pad\'{e} approximation can be used in conjunction with the previous methods to give
a robust framework for analyzing complex singularities. For example, 
one can first trace all possible singularities in the complex plane by
evaluating 
the Pad\'{e} approximation followed by application of the BPH method in order to focus on the
positions where
singularities lie in order to retrieve information on the characterization of the
singularities.
\subsection{Singularity analysis: Prandtl results}
\label{Wssa}

In the remainder of Section~\ref{SA}, analysis of the complex singularities of
the wall shear stress is performed 
for both Prandtl and Navier-Stokes cases.
It will be shown how at $t_s\approx1.5$ a singularity forms in Prandtl's wall
shear.
Moreover, the possible links between the various stages
of the unsteady separation process will be investigated,
and the characterization of the complex singularities of the wall shear in Navier-Stokes
solutions will be accomplished. In particular, it will be shown that a complex singularity, which can be
classified
as the same kind as the VDS singularity in Prandtl's equations, is also present in Navier-Stokes,
and the large- and small-scale interactions can be related to two distinct
groups of complex singularities.

\begin{figure}
\includegraphics[width=11cm]{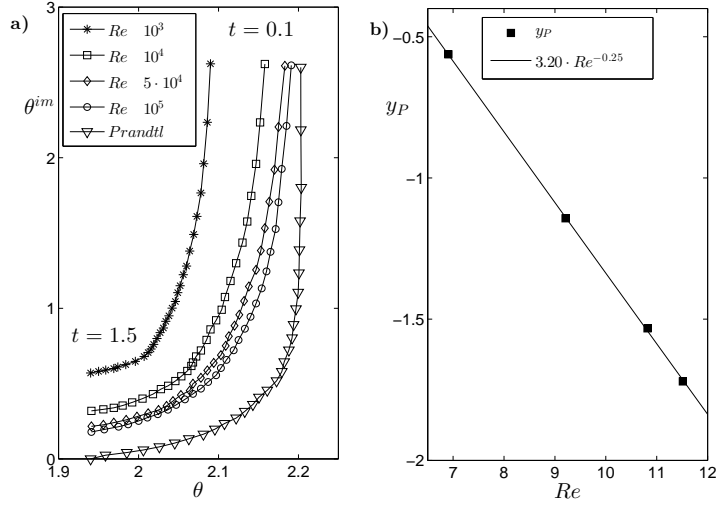}
\caption{\textbf{a)} The temporal evolutions in the complex plane
$(\theta,\theta^{im})$ of the complex singularity of $\tau_w^P$ and the complex
singularities $s_P$ of $\tau_w^{NS}$ for the various Reynolds numbers from time 0.1 up to time 1.5 with
temporal step of 0.05. At $t_s=1.5$, the singularity of $\tau_w^P$ hits the real
axis at $\theta\approx1.94$, while for $\tau_w^{NS}$ the singularity remains at a
distance $y_P$ from the real axis, which goes like $3.2\cdot Re^{-0.25}$.\textbf{ b)} 
The distance $y_P$ is shown versus the Reynolds numbers in log-log coordinates. }
\label{sing_PRA_ALL_2}
\end{figure}

\begin{figure}
\subfigure[Fourier spectrum $\tau_k^P$ of Prandtl's wall shear at various times.]{\includegraphics[width=6.5cm]{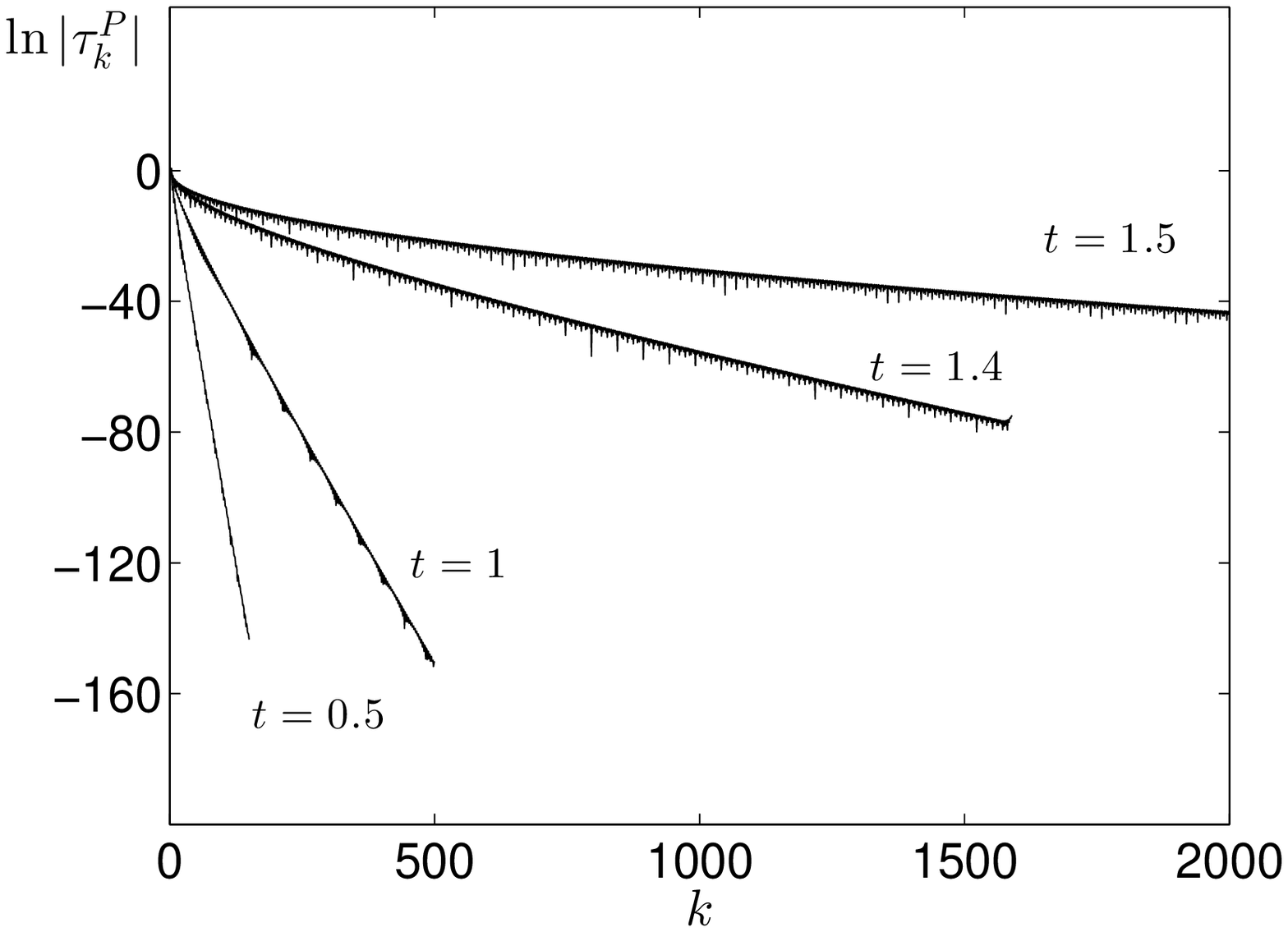}}
\subfigure[The indicatrix function $h(x)$.]{\includegraphics[height=4.8cm,width=6.5cm]{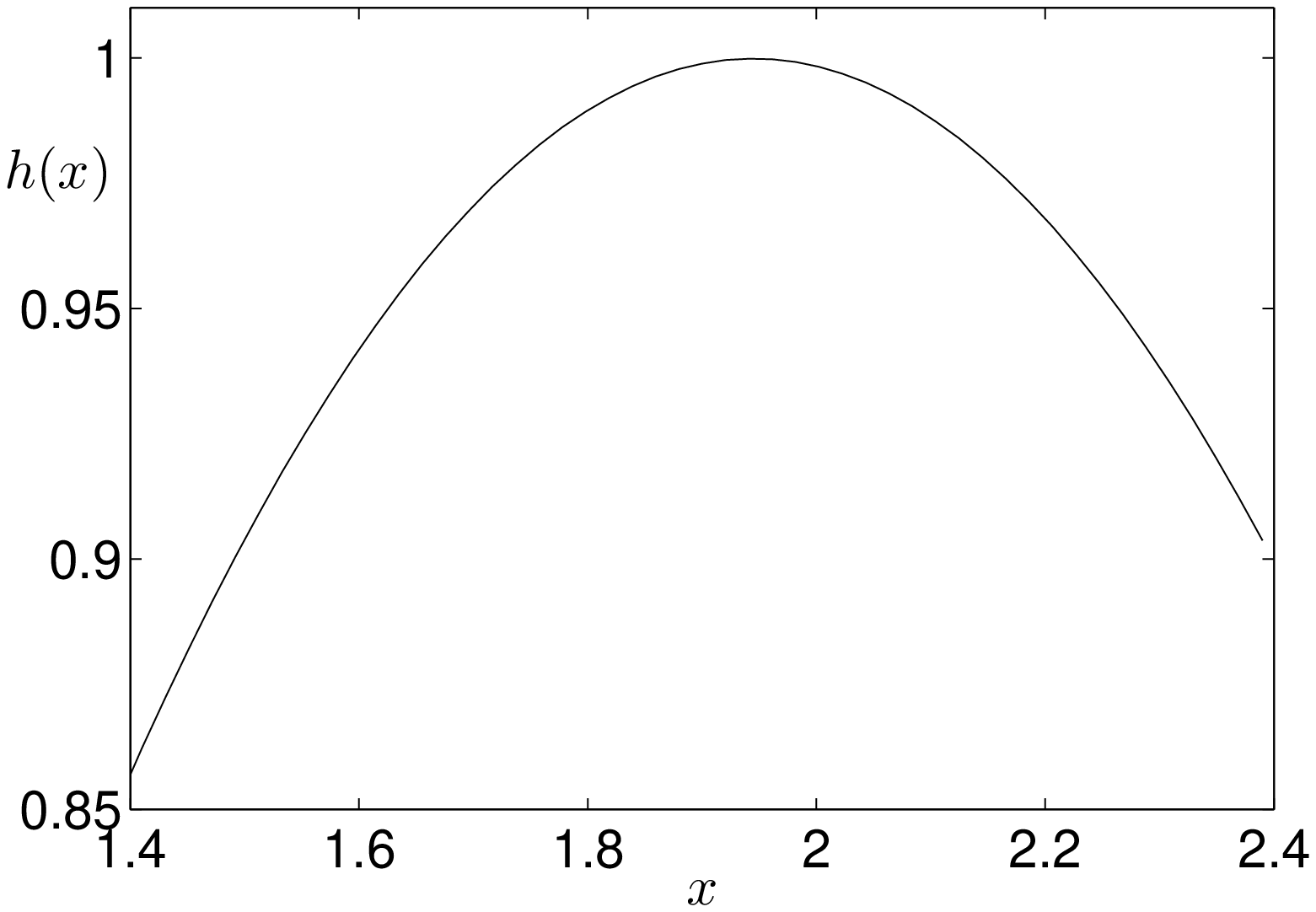}}
\caption{\textbf{a)} The Fourier spectrum $\tau_k^P$ of Prandtl's wall shear at various times.  At $t_s=1.5$ the spectrum totally loses exponential 
decay indicating formation of a singularity. \textbf{b)} The indicatrix function $h(x)$ evaluated through the BPH method at
$t_s=1.5$. $h$ behaves like a cosine function of amplitude one centered at $x_s\approx1.94$, which again is indicative of singularity formation.}
\label{borel_prandtl_cos}
\end{figure}

\begin{figure}
\subfigure[The Fourier spectrum $\tau_k^P$ for Prandtl's wall shear.]{\includegraphics[width=6.5cm]{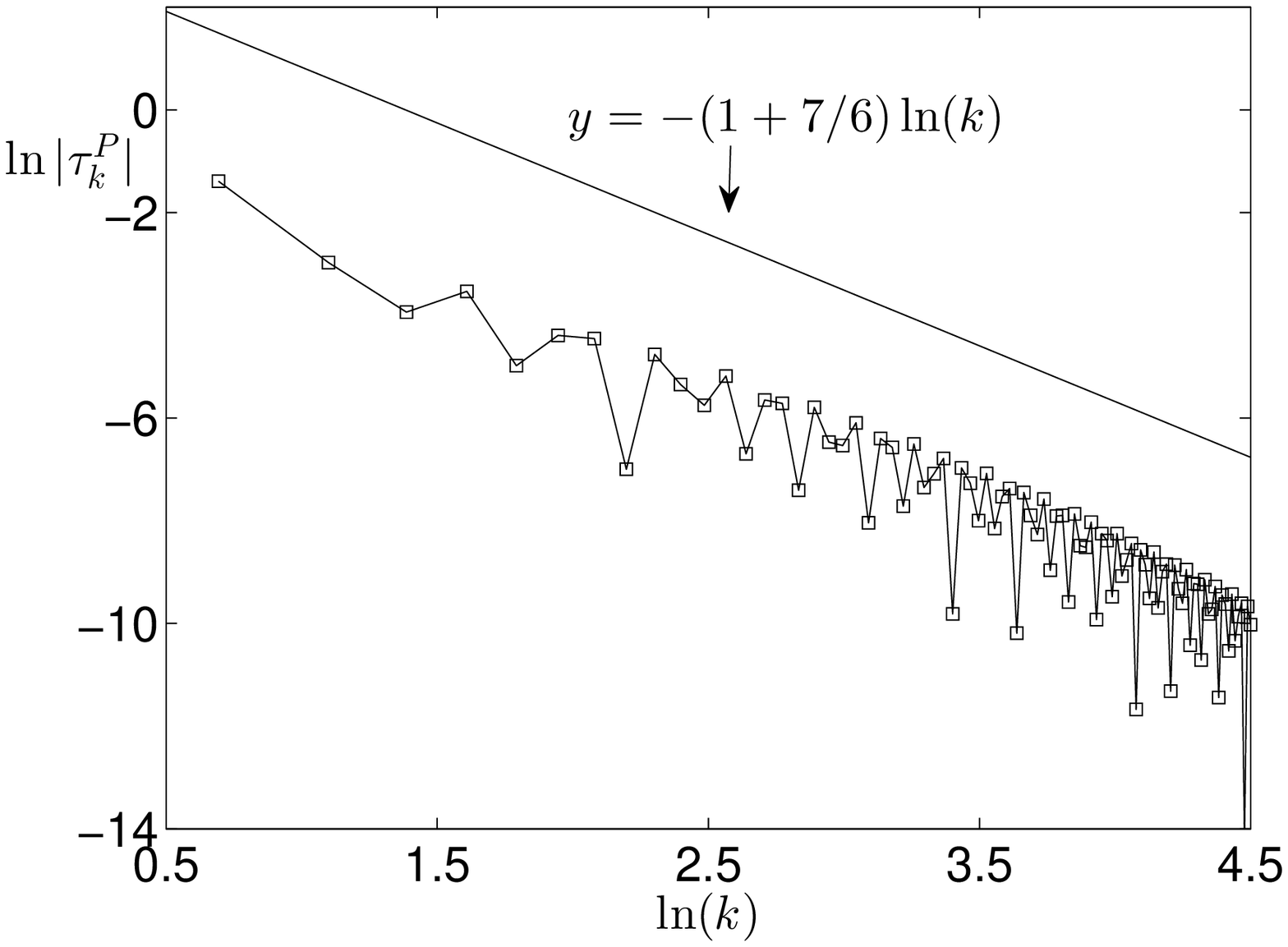}}
\subfigure[The rate of algebraic decay $\alpha^P$.]{\includegraphics[width=6.5cm]{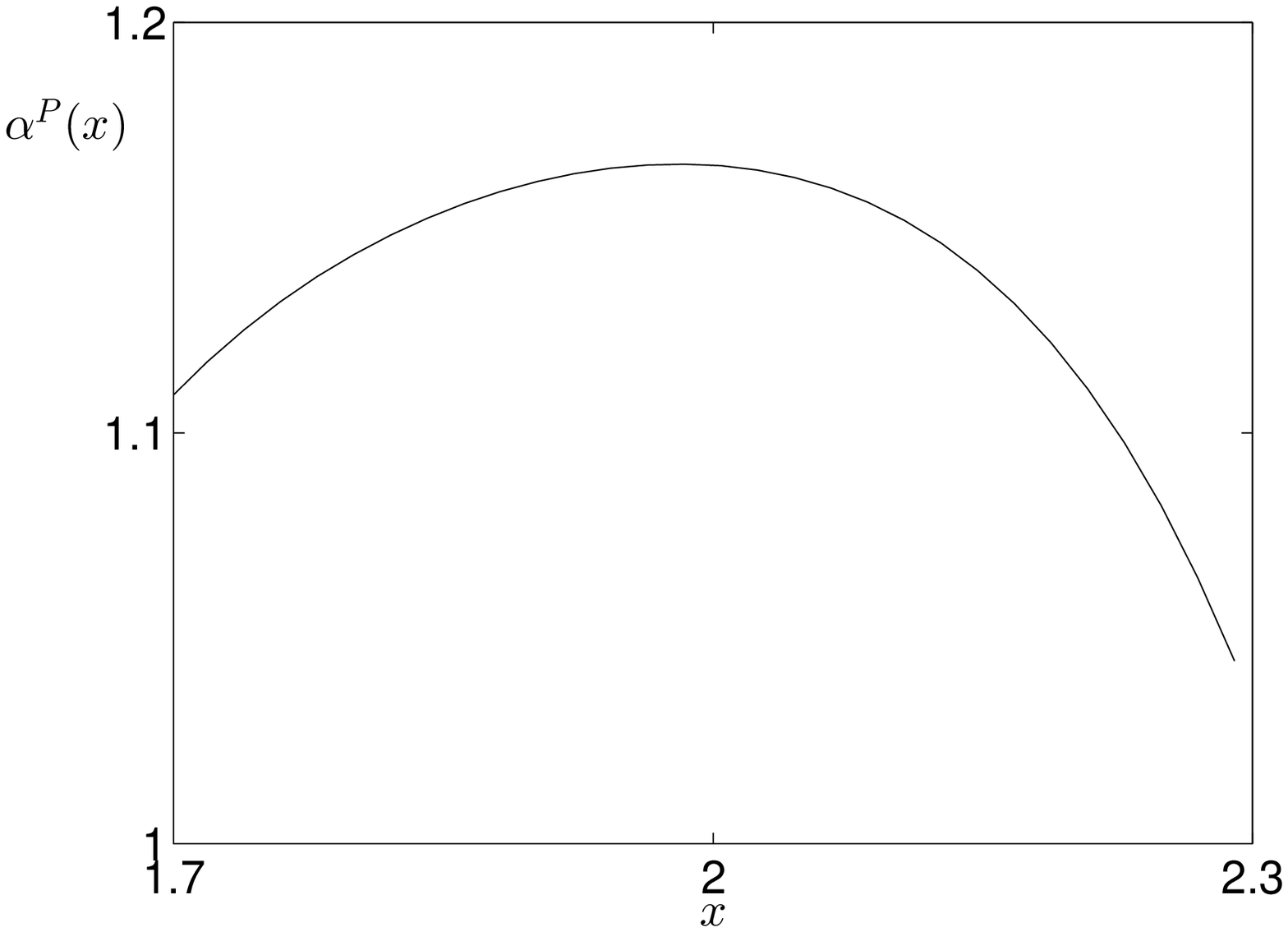}}
\caption{The characterization $\alpha^P$ of the singularity in $\tau_w^P$ at
$t_s=1.5$.  \textbf{a)} The Fourier spectrum of $\tau_k^P$ for Prandtl's wall shear at $t_s=1.5$
in log-log coordinates.  Its slope agrees
with a straight line of slope $-(1+7/6)$, meaning that the characterization of the
singularity is $\alpha_P=7/6$.
\textbf{b)} The rate of algebraic decay $\alpha$ evaluated from equation \eqref{borel_as} at $t_s=1.5$.  The value at $x_s=1.94$, where  the singularity forms, is $\alpha \approx7/6$.}
\label{borel_prandtl_alpha}
\end{figure}

\begin{figure}
\begin{center}
\vspace*{-0cm}\includegraphics[width=8.5cm]{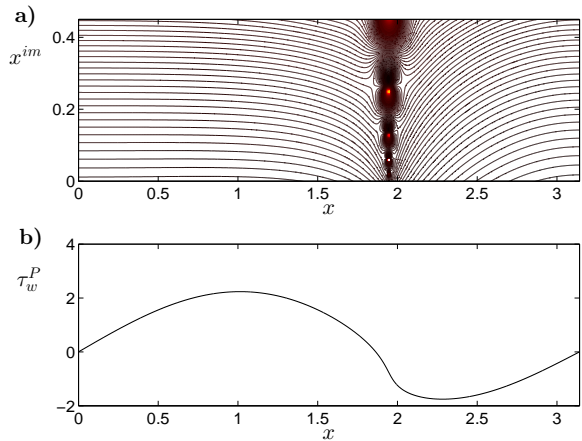}
\caption{The contour levels of the modulus of the Pad\'{e} approximant $P_{200/200}$ of
$\tau_w^P$ at $t=1.495$, with the singularity being very close to the real axis.  Because 
the singularity is a branch cut, the Pad\'{e} approximant can only approximate the branch cut as a series
of poles along where the branch should be. \textbf{b)} $\tau_w^P$ at
$t=1.495$.}
\label{pra_pade_ws1p5}
\end{center}
\end{figure}

As stated in Section~\ref{PS}, a singularity forms at time $t_s$ in Prandtl's solution.
In \cite{GSS09} it was shown that this singularity is manifest as a
shock in the velocity component $u(x^*=1.94,Y^*\approx7)$.
Moreover, through the singularity-tracking method described in Section
\ref{sing_methods},
this singularity has been classified as a cubic-root singularity.
At $t_s$, Prandtl's wall shear $\tau_w^P$ is also characterized by formation of a
singularity. In fact, applying the methods outlined in the previous
section, the temporal evolution of the singularity in
$\tau_w^P$ in the complex plane has been tracked up to time
$t_s$. This evolution is shown in Figure~\ref{sing_PRA_ALL_2} from $t=0.1$ to
$t=1.5$ with a time step of $0.05$.
One can see that up to the time $t_r\approx0.35$, when the recirculation region
forms, the
singularity approaches the real axis along a curve with nearly constant real part.  Then the real part of the complex
singularity moves
toward the position $x_s=1.94$, and at time $t_s \approx 1.5$, the singularity hits
the real axis close to the point of zero wall shear. It is clear, therefore, 
that the physical meaning that is
attributable to this singularity is the formation of the recirculation region.
In Figure~\ref{borel_prandtl_cos}a,b, the Fourier spectrum of the wall shear
and the indicatrix function $h$ obtained from \eqref{borel_as}
are shown at $t=t_s$.  
The singularity formation is revealed by the total loss
of exponential decay in the spectrum ($\delta=0$)
and by the indicatrix function $h$, which is represented by a cosine function of
amplitude one centered at $x_s \approx 1.94$.
The characterization of this singularity has been investigated by evaluating 
the rate of algebraic decay in \eqref{asyspectrum} and \eqref{borel_as}, and we
have obtained the common value $\alpha\approx7/6$, which reveals that the wall
shear blows up in the second derivative. In Figure~\ref{borel_prandtl_alpha}a,
the Fourier spectrum
in log-log coordinates is shown at $t=t_s$, and its slope agrees closely with
a straight line of slope $7/6+1$. 
In Figure~\ref{borel_prandtl_alpha}b, the
rate of algebraic decay $\alpha^P(x)$ from \eqref{borel_as} is shown at $t=t_s$,
and one can see that $\alpha(x=1.94)\approx7/6$.
We conclude this section by showing in Figure~\ref{pra_pade_ws1p5} 
the modulus of the Pad\'{e} approximant $P_{200/200}$ of the wall shear at
$t_s=1.5$ in the complex plane.  
It would be expected that a branch cut appears along the line
passing through $x_s = 1.94$ and parallel to the imaginary axis, but it is well
known (see \cite{BGM96}) that the Pad\'{e} approximant 
approximates a branch cut as a series of poles collapsing where the branch should
be, as is shown in Figure~\ref{pra_pade_ws1p5}.

\subsection{Singularity analysis: Navier-Stokes results}
\label{NS}

\begin{figure}
\begin{center}
\includegraphics[width=10.5cm]{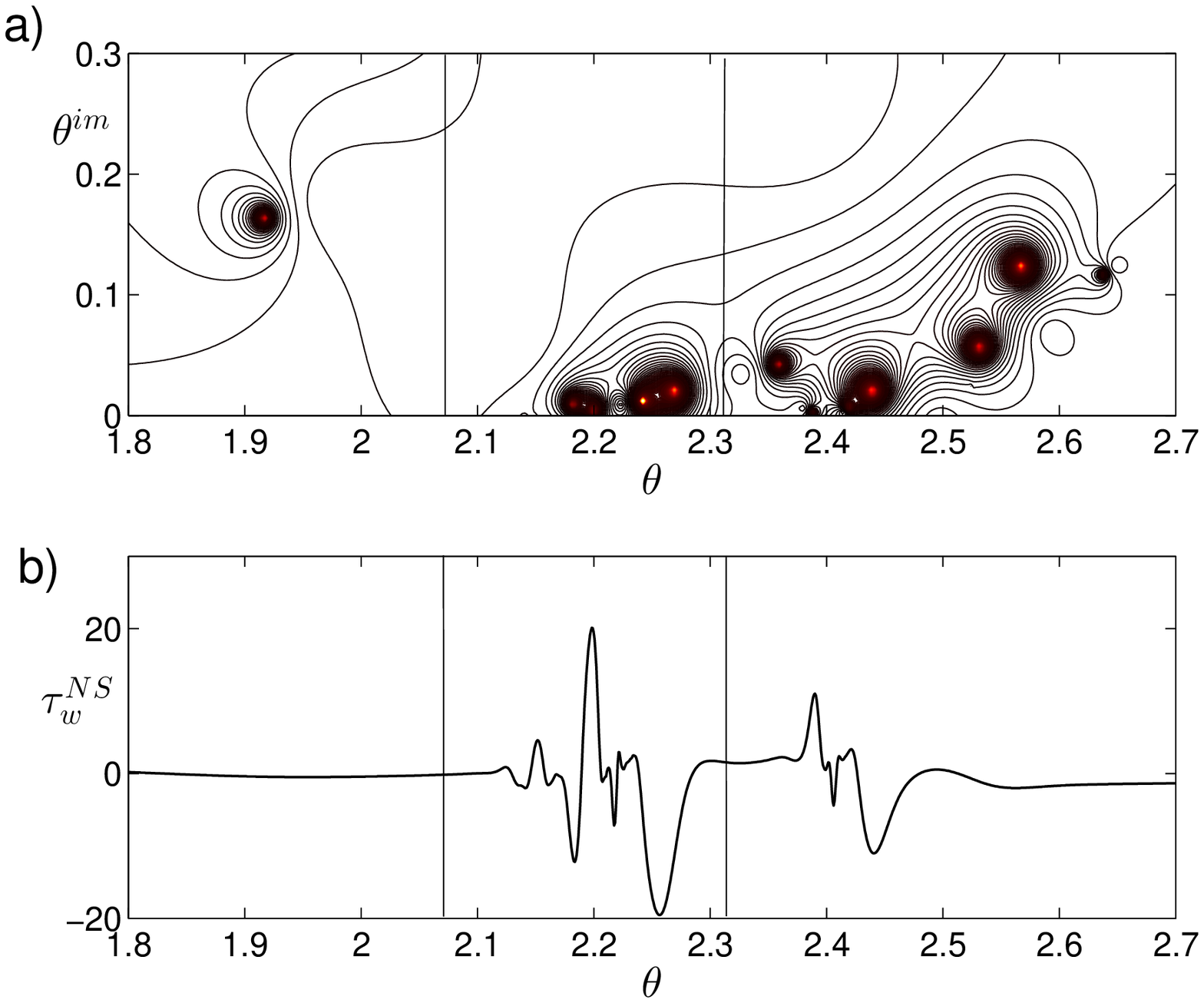}
\caption{\textbf{a)}  The contour levels of the modulus of the Pad\'{e} approximant $P_{300/300}$ of
$\tau_w^{NS}$ for $Re=10^5$ at $t=1.58$.  Three distinct groups of
complex singularities are present. In the left group only the singularity $s_P$
is present, and it corresponds to the wall shear singularity in Prandtl's equations. In the middle group
there are several singularities that correspond to the large-scale interaction. The right group consists of singularities that correspond to the small-scale interaction. The latter group is present only for
moderate to high Reynolds numbers ($Re\geq10^4$).
\textbf{b)} The wall shear $\tau_w^{NS}$, where the singularities in the three groups
correspond to the high gradients forming in $\tau_w^{NS}$.}
\label{singgroup}
\end{center}
\end{figure}

As shown in Section \ref{NSS}, the wall shear $\tau_w^{NS}$ is an
indicator
revealing the onset of the various stages of the separation process in Navier-Stokes solutions.
In fact, the first relevant viscous-inviscid interaction visible in Navier-Stokes solutions, i.e.\ large-scale
interaction, leads to the first relevant
quantitative differences between the Navier-Stokes and Prandtl wall shear. These differences
become more evident
during the small-scale interaction stage for $ Re \geq O(10^4)$, and it is of primary interest in the present 
investigation to find the relationship between these differences
and the presence of complex singularities in $\tau_w^{NS}$.
It will be shown that $\tau_w^{NS}$ has several singularities that can be divided
into three distinct groups.  These three groups of singularities are visible in Figure~\ref{singgroup}, which shows the modulus of the Pad\'{e} approximant $P_{300/3000}$ of the wall shear for
$Re=10^5$ at $t=1.58$, which is well after the onset of small-scale interaction.  
The first group of singularities is present for every Reynolds number, and it consists of only
one singularity that is indicative of the singularity in $\tau_w^P$.  The second
group of complex singularities is still present for each Reynolds number, and it is
related to the large-scale interaction.  The third group of singularities is only
present for $Re \geq O(10^4)$ and characterizes the small-scale interaction.

\subsubsection{Singularity analysis: van Dommelen \& Shen's singularity}

Let us now discuss the physical phenomena that can be related to these groups of singularities.
The first singularity in $\tau_w^{NS}$ is comparable with the singularity of $\tau_w^P$,
and we shall call this singularity $s_{P}$. We have tracked in time the
position of $s_P$ in the complex plane through the singularity-tracking methods,
and this temporal motion is shown in Figure~\ref{sing_PRA_ALL_2}a for each Reynolds number 
from $t = 0.1$ to time $t_s = 1.5$ with temporal step of $0.05$.
The qualitative behavior of this singularity
is similar for each Reynolds number and matches closely that observed in the Prandtl case.  The
singularities rapidly move toward the real axis slightly shifting along the
angular direction $\theta$ upstream on the circular cylinder. This reflects the physical fact that the recirculation
region attached to the circular cylinder increases its size along the angular (streamwise) direction. As
previously observed, the singularity of $\tau_w^P$
gets very close to the point of zero wall shear stress,
which moves upstream on the cylinder surface.  Therefore, it is expected that the location of the real
part of the singularity $s_P$
also moves upstream on the circular cylinder following the location of the zero
wall-shear point.
At time  $t_s$, all the singularities have a real
position close to $x_s\approx1.94$ (where VDS singularity forms), but at a
distance $y_P$ which follows the relationship 
$y_{P}=C_{P}Re^{\lambda_P}$, where $\lambda_P\approx-0.25$ and $C_P\approx3.2$
(see in Figure~\ref{sing_PRA_ALL_2}b, where $y_{P}$ is shown versus the
Reynolds number in log-log
coordinates).

\begin{figure}
\begin{center}
\includegraphics[width=10.5cm]{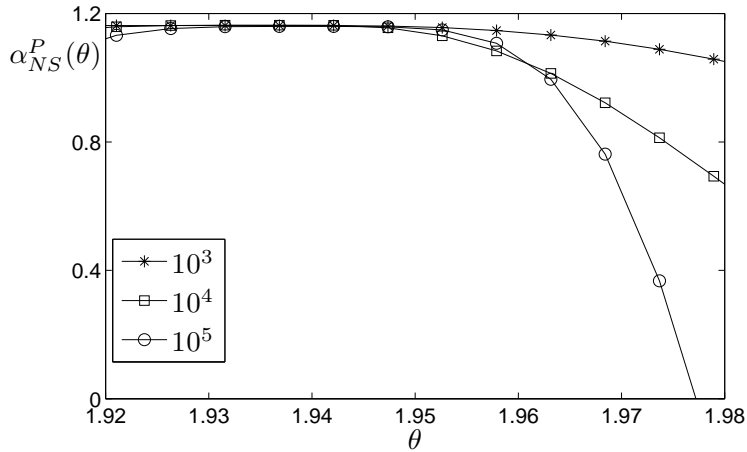}
\caption{The characterization $\alpha_{NS}^P$ of the complex singularity $s_P$
of $\tau_w^{NS}$ evaluated through the BPH method.  At $t_s=1.5$, the location of the real part of the singularity is $\theta \approx1.94$ and
$\alpha_{NS}^P\approx7/6$ for $Re=10^3,10^4,10^5$. }
\label{borel_NS_ALPHA_P}
\end{center}
\end{figure}

The primary similarity between $s_{P}$ and the singularity that occurs in
$\tau_{w}^P$ lies in
their characterization.  
It has been determined through the BPH method that close to the time of
singularity formation for Prandtl's equations, the algebraic characterization
of $s_{P}$ is 
$\alpha_{NS}^P\approx7/6$ for each Reynolds number (see Figure~\ref{borel_NS_ALPHA_P}
where $\alpha_{NS}^P$ is shown at time $t_s=1.5$ for $Re=10^3,10^4,10^5$). As
compared
to the Prandtl case, the characterization of $\alpha_{NS}^P$ has been more difficult
to evaluate because the function \eqref{borel_as} is more difficult to handle numerically.  
This is due to the various complex singularities (introduced in the following
sections) that affect the indicatrix function.

\subsubsection{Singularity analysis: large-scale interaction singularities}
\label{lssing}

The second group of complex singularities in $\tau_w^{NS}$ exists for each
Reynolds number, and this group is related to the large-scale interaction.
These singularities are always located downstream of the 
singularity discussed in the previous section (see
Figure~\ref{singgroup}), and they are very close to each other especially for
higher Reynolds number. This proximity makes it extremely difficult to precisely
characterize these singularities using the numerical methods described in Section
\ref{sing_methods}.
The most accurately resolved singularity in this group is the one closest to the real axis for all time; we shall call this singularity $s_{ls}$.  The remaining singularities
in this group are difficult to distinguish using the BPH 
method.  Although it does not provide as much information about their characterization, Pad\'{e} approximants have been much more useful in tracking their position. For the kind
of analysis to be performed here, however, it is enough to recover information only on
$s_{ls}$ in order to characterize the large-scale interaction stage.

\begin{figure}
\begin{center}
\includegraphics[width=13.5cm]{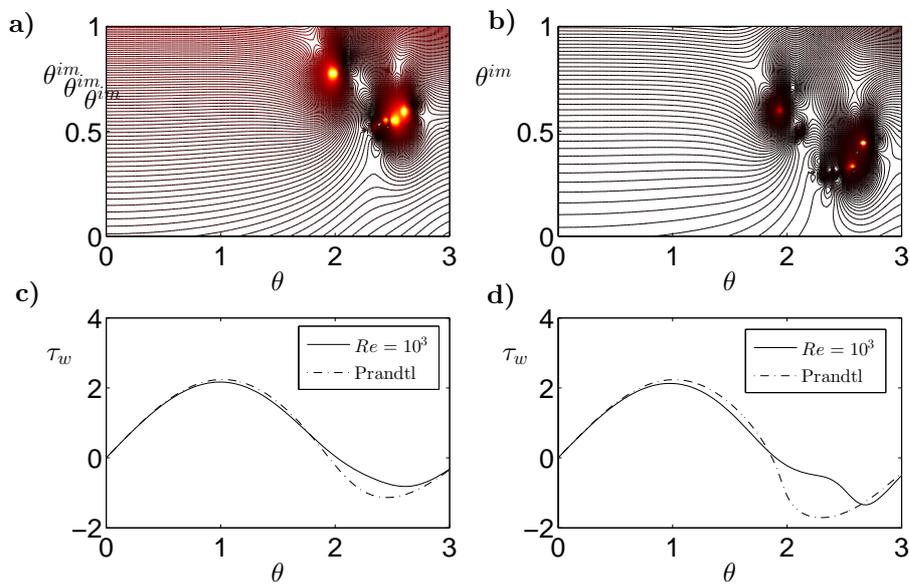}
\caption{ The comparison between the wall shear $\tau_w$ for Navier-Stokes and Prandtl cases,
and the modulus of Pad\'{e} approximants $P_{200/200}$ of $\tau_w$ for Navier-Stokes for $Re=10^3$
at $t=1$ (on the left) and $t=1.45$ (on the right). The singularity $s_{ls}$ is located at $2.5+0.55i$
and $2.52+0.33i$ at $t=1,1.45$, respectively. At $t=1.45$, $s_{ls}$ corresponds to formation of the gradient in
$\tau_w$ close its minimum.}
\label{padels_10p3}
\end{center}
\end{figure}

\begin{figure}
\begin{center}
\subfigure[$Re=10^3, t=2.3$]{\includegraphics[width=7.5cm]{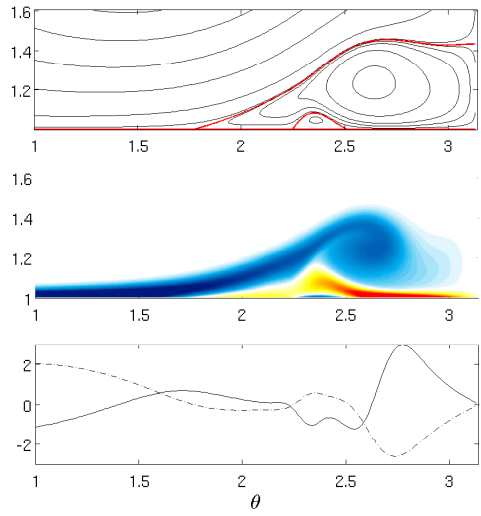}}
\subfigure[$Re=10^5, t=1.45$]{\includegraphics[width=6.75cm]{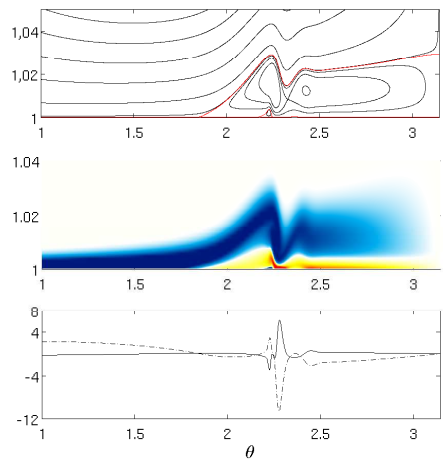}}
\caption{ Upper figures: the streamlines (zero level in red). Middle figures:
vorticity contour levels (blue colours negative vorticity,
red colours   positive vorticity). Lower figures: $\partial_{\theta}p_w$ (solid)
and  $\tau_w$ (dashed).}
\label{RE10p3ls}
\end{center}
\end{figure}

To show how $s_{ls}$ is related to large-scale interaction, let us focus on
the case with $Re=10^3$.
As pointed out in Section \ref{NSS}, the large-scale interaction
begins to strongly influence the flow evolution when $\partial_{\theta}p_{|r=1}$
and $\tau_{w}^{NS}$ quantitatively differ from the same quantities of Prandtl's
solution.
In Figure~\ref{padels_10p3}, the wall shear is shown for
$Re=10^3$ at $t=1$ (when large-scale interaction has just begun) and $t=1.45$,
and compared to the contour levels of the modulus of Pad\'{e} approximants $P_{200/200}$ of
$\tau_{w}^{NS}$ in the complex semi-plane.  
The singularities, which are visible as poles, are
located where the contour levels become most dense.  At $t=1$, the singularity $s_{ls}$ is located at
$2.5+0.55i$, and it clearly corresponds
to the variation in the wall shear close to its local minimum as compared to
Prandtl's wall shear.  At $t=1.45$, the singularity $s_{ls}$
gets closer to the real axis, and it is located at $2.52+0.33i$, and this leads
to a more dramatic change in the wall shear.  In fact, a gradient with respect to the angular
coordinate $\theta$ forms in correspondence to $s_{ls}$. As time passes 
this gradient becomes stronger, and a pair of
positive-negative critical points form in the wall shear (see for example
Figure~\ref{RE10p3ls}a, where the pair of positive-negative critical points is visible close to
$\theta=2.6$ for $Re=10^3$ at $t=2.3$). 
This means that a new recirculation region forms that is attached to the circular cylinder beneath the primary
recirculation region. This new recirculation region, however, is not related to
the formation of the kinks in the streamlines and vorticity that characterize
the small-scale interaction stage.  This explains
how this group of singularities, led by $s_{ls}$, does not correspond to the
small-scale interaction stage. 
%

\begin{figure}
\begin{center}
\includegraphics[width=11.5cm]{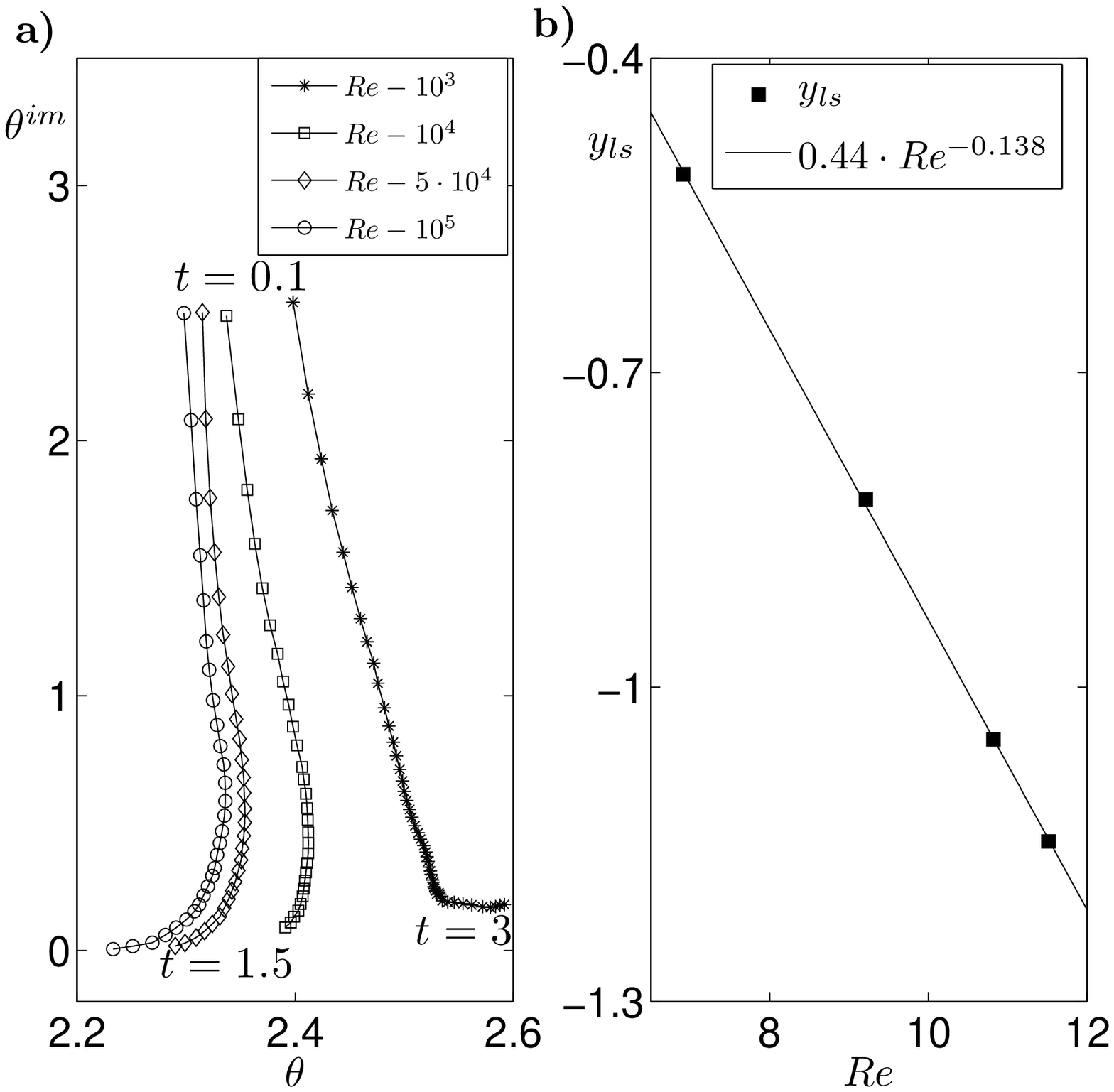}
\caption{\textbf{a) }The temporal evolution in the complex plane
$(\theta,\theta^{im})$ of the complex singularity $s_{ls}$
of $\tau_w^{NS}$ for $Re=10^3$ from time 0.1 up to time 3 with temporal step of
0.05 and for $Re=10^4,5\cdot 10^4,10^5$ 
from time 0.1 up to time 1.5 with temporal step of 0.05. After large-scale
interaction begins, the location of the real part of $s_{ls}$
moves upstream along the circular cylinder for $Re=10^4-10^5$, while for $Re=10^3$ the real
location of $s_{ls}$ moves downstream along the
circular cylinder even after total detachment of the boundary layer.  \textbf{b)}  $y_{ls}$ is shown versus the Reynolds number in
log-log coordinates at the
time $T_{LS}$ at which large-scale interaction begins 
(see Table~2). The singularity is at a distance $y_{ls}$ from the real
axis that goes like $0.44\cdot Re^{-0.138}$. }
\label{sls}
\end{center}
\end{figure}

The temporal evolution of the complex position of $s_{ls}$ is shown in Figure~\ref{sls}a for all
the Reynolds numbers considered from time $t=0.1$ up to time $t=1.5$ for $Re=10^4,5\cdot10^4,10^5$
and from time $t=0.1$ up to time $t=3$ for $Re=10^3$. 
This temporal evolution is quite similar for all $Re>10^3$, while for $Re=10^3$ the
evolution of $s_{ls}$ exhibits a distinctly different behavior. The most relevant difference arises during the large-scale
interaction stage.  For $Re>10^3$, for example, $s_{ls}$ changes its motion by shifting upstream along the circular cylinder, while for $Re=10^3$ it continues to shift downstream along the circular cylinder even after total
detachment of the boundary layer. This distinct motion of $s_{ls}$ for $Re=10^3$ 
can be explained by the fact that no small-scale interaction develops after the
large-scale interaction, and no strong gradients forms in $\tau_w^{NS}$.
These gradients forming for cases with $Re\geq10^4$ are due to the presence
of a third group of singularities as we shall see in the following subsection.  It has been observed that at the time
at which large-scale interaction begins, the distance $y_{ls}$ from the real axis to the singularity $s_{ls}$ follows
the relationship $y_{ls}=C_{ls}Re^{\lambda_{ls}}$, where $\lambda_{ls}\approx-0.138$ and
$C_{ls}\approx0.447$, as one can see in Figure~\ref{sls}b, where $y_{ls}$ is shown versus Reynolds number in
log-log coordinates.

\begin{figure}
\begin{center}
\includegraphics[width=10.5cm]{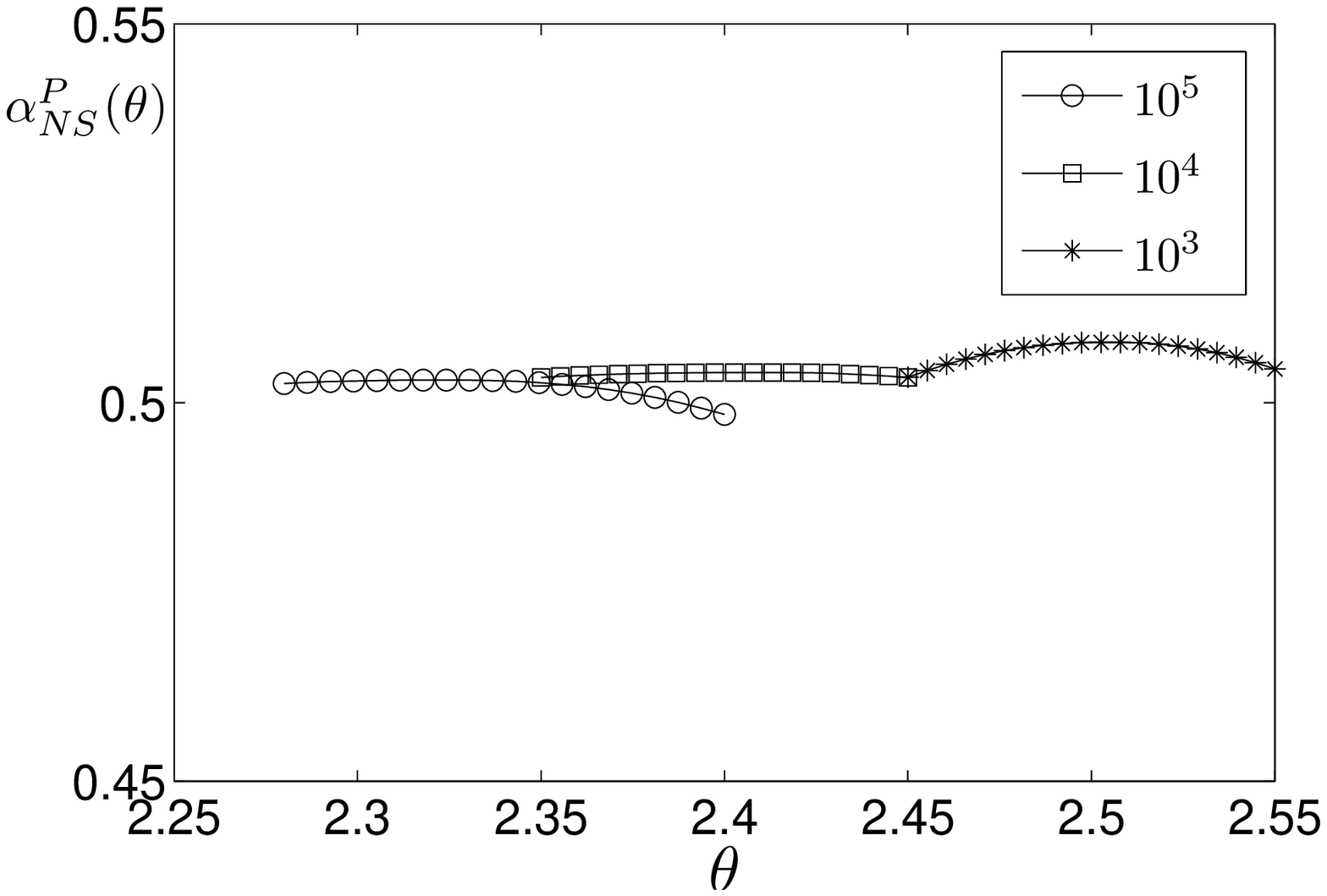}
\caption{The characterization $\alpha_{NS}^{s_{ls}}$ of the complex singularity
$s_{ls}$ of $\tau_w^{NS}$ evaluated from the BPH method.  At $t=1$, $\alpha_{NS}^{s_{ls}}\approx0.5$ for $Re=10^3,10^4,10^5$, and $s_{ls}$ is located at $(2.50,0.55),(2.41,0.38),(2.32,0.29)$, respectively.}
\label{borel_NS_ALPHA_ls}
\end{center}
\end{figure}

Based on the BPH method, we have determined that the characterization of the
singularity $s_{ls}$ is $\alpha_{NS}^{s_{ls}}\approx0.5$ for all the Reynolds numbers considered.  This value is
consistent with the formation of the
gradient in the wall shear.  The characterization $\alpha_{s_{ls}}$ is accurately determined between $t=0.9$ and $t=-1.2$
when the other complex singularities are still far enough away from the real axis so as not to interfere with the singularity 
of interest.
In Figure~\ref{borel_NS_ALPHA_ls}, the rate of algebraic decay
$\alpha_{NS}^{s_{ls}}(\phi)$ is shown 
from equation \eqref{borel_as} for $t=1$ for the various Reynolds numbers, and one can see that for the Reynolds numbers considered, $\alpha_{NS}^{s_{ls}}(\phi)\approx0.5$.

\subsubsection{Singularity analysis: small-scale interaction singularities}
\label{sssing}

\begin{figure}
\begin{center}
\includegraphics[width=11.5cm]{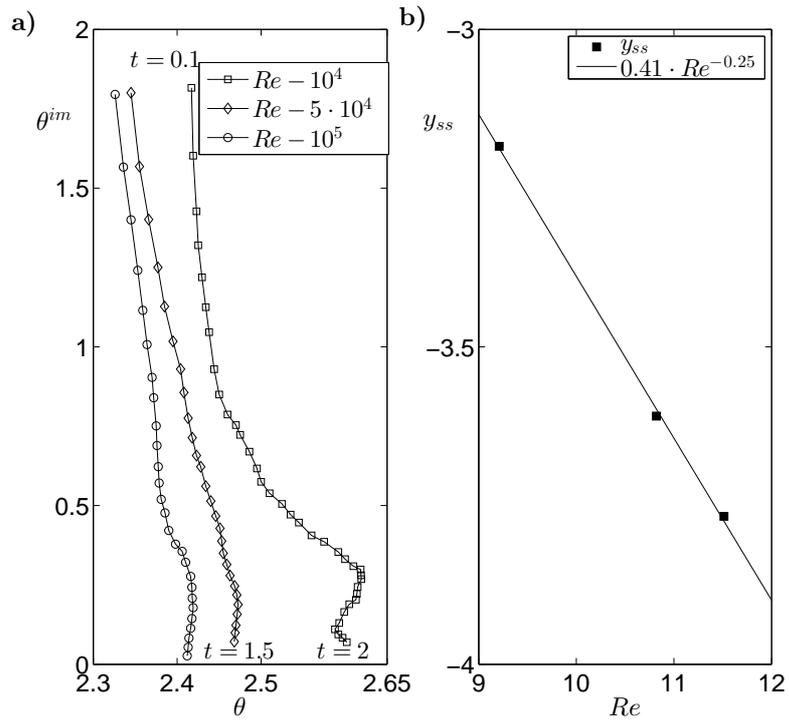}
\caption{\textbf{a)} The temporal evolution in the complex plane
$(\theta,\theta^{im})$ of the complex singularity $s_{ss}$
of $\tau_w^{NS}$ for $Re=10^4$ from time 0.1 up to time 2 with temporal step of
0.05 and for $Re=5 \cdot 10^4,10^5$ from time 0.1 up to time 1.5 with 
temporal step of 0.05.\textbf{ b)} $y_{ss}$ is shown versus the Reynolds number in
log-log coordinates at the onset of small-scale interaction (see
Table~2). The singularity is at a distance from the real
axis that goes like $0.41\cdot Re^{-0.25}$.}
\label{sss}
\end{center}
\end{figure}

As described in Section \ref{NSS}, the onset of small-scale interaction is
characterized by the formation of a kink in the streamlines and vorticity
contours, splitting of the recirculation region, and formation of
strong gradients in $\tau_w^{NS}$ and
$\partial_{\theta}p_w$. All of these phenomena are visible in
Figure~\ref{RE10p3ls} for $Re=10^5$ at $t=1.45$.
To understand how the complex singularities in $\tau_w^{NS}$ are related to the
small-scale interaction, it is first observed that the formation of large
gradients in $\tau_w^{NS}$ below the
zone of the primary recirculation region are clearly characterized by the
presence of
a third
group of complex singularities. As in the case of the group of singularities
characterizing the large-scale interaction, this group has a primary singularity that is always closest to
the real axis. This singularity, which we shall call $s_{ss}$, is the one
corresponding to the
gradient close to $\theta\approx2.45$ on the right of the maximum of
$\tau_w^{NS}$
visible in Figure~\ref{streamDISK105p4}b for $Re=10^5$ at $t=1.4$.
To understand how this singularity behaves according to the $Re$ number, we show in Figure~\ref{sss}a the temporal evolution of the position of $s_{ss}$ in the
complex plane for $Re = 10^4$ from time $t = 0.1$ up to time $t = 2$, and for $Re=5 \cdot 10^4$
and $Re=10^5$ for $t=0.1$ up to $t=1.5$ with time step of $0.05$. 
This evolution is not as smooth as compared to that of
$s_P$ and $s_{ls}$. In the first phase, which also encompasses the large-scale interaction stage,
the singularities tend to get closer to the real axis by
shifting downstream along the circular cylinder. When the effects of the small-scale
interaction begin to be noticeable, however, the singularities
move upstream along the circular cylinder, and this takes place at
$t_{ss}=1.505,1.29,1.26$ at $\theta_{ss}=1.505,1.29,1.26$ for
$Re=10^4,5\cdot10^4,10^5$, respectively.
These times correspond quite well to the times when critical points in
$\partial_{\theta}p_w$ and $\tau_w^{NS}$ form. These critical points represent a minimum and maximum in
$\partial_{\theta} p_w$ and $\tau_w^{NS}$.  In Table~2, 
we report for the various Reynolds numbers the times and the locations 
$t_p$, $t_w$ and $\theta_p$, $\theta_w$ at which these points form, and the
times $t_{ss}$ and the real locations $\theta_{ss}$  
where the singularities $s_{ss}$ change their motion from downstream to upstream
along the circular cylinder.
In particular, the formation of the
maximum in $\tau_w^{NS}$, and the relatively large gradients that form,
is a direct consequence of the singularity $s_{ss}$ getting close to the real
 domain. 
 
In light of the physical events that correspond to the onset of the small-scale interaction, 
a beginning time can be estimated.  As $\partial_{\theta}p_w$ 
has a physical meaning in the flow evolution, i.e.\ it represents the acceleration
along the circular cylinder, and we have used it to define
the beginning of the large-scale interaction in the previous section, we
propose as the time 
of beginning of the small-scale interaction 
the time at which the minimum in $\partial_{\theta}p_w$ appears downstream of the maximum (this minimum does not form for $Re=10^3$).  For example,  
this minimum is visible in Figure~\ref{streamDISK105p4}b at
$\theta\approx2.45$ for $Re=10^5$ at $t=1.4$. The physical meaning is clear; 
when this minimum  becomes negative, the pressure gradient
is adverse with respect to the flow direction between the primary recirculation region
and the cylinder surface. 
The fluid between the minimum and the maximum is therefore strongly
compressed in the streamwise direction, accelerating the evolution of the kink in
the streamlines leading to a 
spike with the consequent splitting of the primary recirculation region.
We have also checked that the formation of this minimum in $\partial_{\theta}p_w$ is
related to the small-scale interaction for other initial conditions, such as the 
thick-core vortex \cite{Cas00,OC02} and the rectilinear vortex
\cite{GSS11}, supporting the 
strong relevance of this event in the flow evolution. As pointed out in Section~\ref{NSS}, the large-scale 
interaction rapidly evolves toward the small-scale interaction as the Reynolds number increases.  

At the time at which small-scale interaction begins, it has been observed that the
distance $y_{ss}$ from the real axis 
of the singularity $s_{ss}$ follows the behavior $y_{ss}=C_{ss}Re^{\lambda_{ss}}$, where
$\lambda_{ss}\approx-0.25$ and $C_{ss}\approx0.41$.  This can be seen in Figure~\ref{sss}b, 
where $y_{ss}$ is shown versus the Reynolds number on log-log
coordinates.

As compared to $s_P$ and $s_{ls}$, the fitting procedures applied in the tracking
methods lead to spurious results. 
In fact, particularly for $Re=5\cdot10^4,10^5$, when $s_{ss}$ gets very close to the
real axis, the fitting procedures do not give reasonable results.
The situation is more clear close to the time at which small-scale interaction
forms, when $s_{ss}$ is far enough from the real axis and from the other
singularities.  In this case, we have obtained a value $\alpha^{ss}_{NS}\approx0.5$ as the
most likely result. This characterization is compatible with the kind of
gradient that forms in $\tau_w^{NS}$ as it clearly shows a growth in the first derivative.

\begin{figure}
\begin{center}
\includegraphics[width=8.5cm]{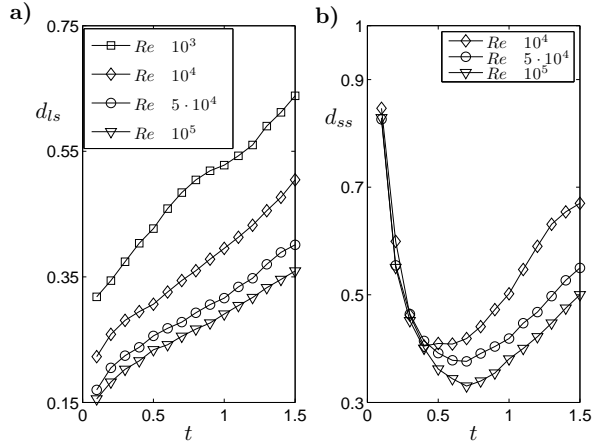}
\caption{\textbf{a)} The temporal evolution of the distance $d_{ls}$ in the
complex plane between the singularities $s_{P}$ and $s_{ls}$ for various Reynolds numbers.
\textbf{b)} The temporal evolution of the distance $d_{ls}$ in the complex plane
between the singularities $s_{P}$ and $s_{ls}$. In both cases, at any fixed time, the distance diminishes as Reynolds number increases, and this supports the
conjecture that asymptotically all the singularities collapse to the
singularity $s_P$. }
\label{distsing}
\end{center}
\end{figure}

The validity of the boundary-layer approximation, which would seem to be questionable from a physical point of view in light of the presence of the large- and small-scale interactions, can be 
deeply corroborated through the present investigation carried out on the complex
singularities of the wall shear. In fact, 
as supposed in \cite{Cas00,OC02} and later in \cite{GSS11}, as
$Re\rightarrow\infty$, large- and small-scale interactions appear to merge into a single interaction that resembles the viscous-inviscid interaction provoked by 
formation of the VDS singularity in Prandtl's equations. 
This is supported by two facts: 1) as Reynolds number increases,
the temporal gap between the onset of large- and small-scale interaction diminishes, and 2) the time of formation of
large-scale interaction, which is $T_{LS}$, tends toward the time at which viscous-inviscid interaction develops in Prandtl's solution
(this trend is admittedly very slow for the Reynolds numbers
considered).
The conjecture that large- and small-scale interaction are
likely to be the same interactions in the limit as $Re \to \infty$ is also supported by
the fact that the distance between the various main singularities of
$\tau_w^{NS}$ diminishes as Reynolds number increases. The temporal evolution of the
distance in the complex plane between $s_P$ and $s_{ls}$ and between $s_P$ and
$s_{ss}$ are shown in Figure~\ref{distsing}a,b for various Reynolds numbers from $t=1$ up to
$t=1.5$.  
This suggests that as Reynolds number
increases, all of the complex singularities converge toward each other and 
eventually collapse to become a single singularity, i.e.\ VDS singularity
$s_P$.

\section{Bidimensional singularity-tracking method: Navier-Stokes results}
\label{STMNS}

In this section, the singularity-tracking method is extended to a bi-variate function
(see \cite{MBF05,PMFB06,GSS09} for details), and the results once again will be related to the various stages of the unsteady separation process.
Given a periodic function that can be expressed as a Fourier series
\begin{equation}
u(x_1,x_2)=\sum_{k_1,k_2} u_{k_1 k_2} e^{i k_1 x_1} e^{i k_2 x_2},
\label{bi_fourier}
\nonumber
\end{equation}
if one considers those modes $(k_1, k_2)$ such that
$k_1= k \cos \theta$ and $k_2= k \sin \theta $, where $k=|(k_1,k_2)|$,
then the asymptotic behavior of the Fourier coefficients in the Fourier
$\mathbf{k}$-space with $k\rightarrow\infty$ have the following asymptotic behavior:
\begin{equation}
u_{k_1 k_2}\approx  k^{-\left(\alpha(\theta)+1\right)} e^{-\delta(\theta) k}
e^{i k x^*(\theta)} \quad \mbox{where} \quad 
(k_1, k_2)= k( \cos \theta,  \sin \theta ).
\label{Fourier_asymp_theta}
\end{equation}
The width of the analyticity strip $\overline{\delta}$ is the minimum over all
directions $\theta$, i.e.
$\delta^*=\min_{\theta}\delta(\theta)$.

The shell-summed Fourier amplitudes, which are a kind of discrete angle average of the Fourier coefficients,
are defined as
$$
A_K \equiv \sum_{K\leq |(k_1,k_2)| < K+1} \left|u_{k_1 k_2} \right|.
$$
The asymptotic behavior of these amplitudes is 
$$
A_K\approx CK^{-\left(\alpha_{Sh}+1/2\right)} \exp{\left(-\delta_{Sh} K\right)}
\quad \mbox{when}\quad
K\rightarrow\infty,
$$
where $\delta_{Sh}$ gives the width of the analyticity strip, while the algebraic
prefactor $\alpha_{Sh}$ gives information on the nature
of the singularity.
As shown in \cite{PMFB06}, using a steepest descent argument,
if one denotes with $\theta^*$ the angle where $\delta(\theta)$
takes its minimum, i.e.\ $\delta^*=\delta(\theta^*)$,
one has that $\delta_{Sh}=\delta(\theta^*)$ and $\alpha_{Sh}=\alpha(\theta^*)$. 
This methodology has been applied in \cite{MBF05} and \cite{PMFB06}  to investigate  the complex singularity 
in the two-dimensional periodic Euler equation,
and in \cite{GSS09} to perform a complete analysis of the singularity formation
for Prandtl's boundary-layer equations in the case of the impulsively started disk.

Following the same procedure applied in \cite{GSS09} for Prandtl's solution, we analyze the spectrum of the velocity component 
$u(r,\theta)$ of the Navier-Stokes solutions. 
To accomplish this, the normal physical domain is mapped to $\Lambda=[1,2]$ as it has been observed that 
for $r \to 2$, the vorticity remains negligible for all Reynolds numbers during the computational time that is the object of our investigation.  
The grid points $\lambda_i$ in $\Lambda$ are exactly the linear mapping of Gauss-Lobatto points 
$\kappa_i=\cos(i\pi/N)_{i=0,\ldots,N}$ in $\Lambda$. 
This allows one to write the solution in terms of its Chebyshev expansion 
\be
u(\theta,\kappa,t)\thickapprox
\sum\limits_{k=-K/2}^{k=K/2}\sum\limits_{j=0}^{j=N}u_{kj}(t)e^{ik\theta}
T_j(\kappa),
\label{ChebT}
\ee
where $T_j$ are the Chebyshev polynomials of the first kind.
Finally, introducing the variable $\zeta=\arccos \kappa$, the
above expression is written as
\be
u(\theta,\zeta,t)\thickapprox
\sum\limits_{k=-K/2}^{k=K/2}\sum\limits_{j=0}^{j=N}u_{kj}(t)e^{ik\theta}
\cos{(j\zeta)},
\label{ChebCos}
\ee
and the singularity-tracking method is applied on the Fourier coefficients
$u_{kj}$.

\begin{figure}
\begin{center}
\includegraphics[width=11.5cm]{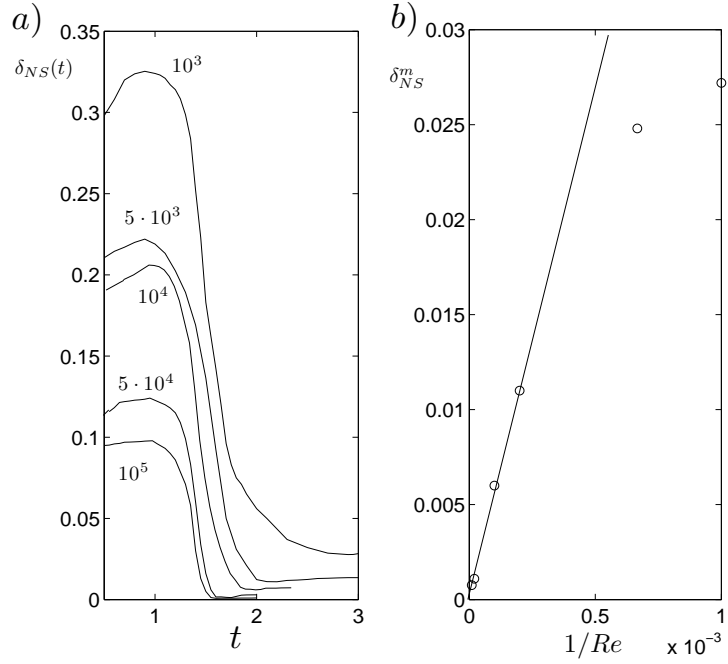}
\caption{\textbf{a)} Temporal evolution of $\delta_{NS}$ for various Reynolds
numbers. After large-scale interaction begins, $\delta_{NS}$ decreases in all
cases owing to the effect of the formation of
strong gradients particularly in the angular direction $\theta$. 
At time $t_s=1.5$, $\delta_{NS}$ approaches zero as Reynolds number increases. Then it
reaches a minimum ($\delta_{NS}^m$) of $O(1/Re)$ in time as shown in \textbf{b)} (at least
for the Reynolds numbers for which small-scale interaction forms). The time at which 
$\delta_{NS}^m$ forms becomes closer to $t_s$ as Reynolds number increases (see Figure~\ref{large_spectrum}b).}
\label{delta_NS_inset}
\end{center}
\end{figure}

\begin{figure}
\begin{center}
\includegraphics[width=9.5cm]{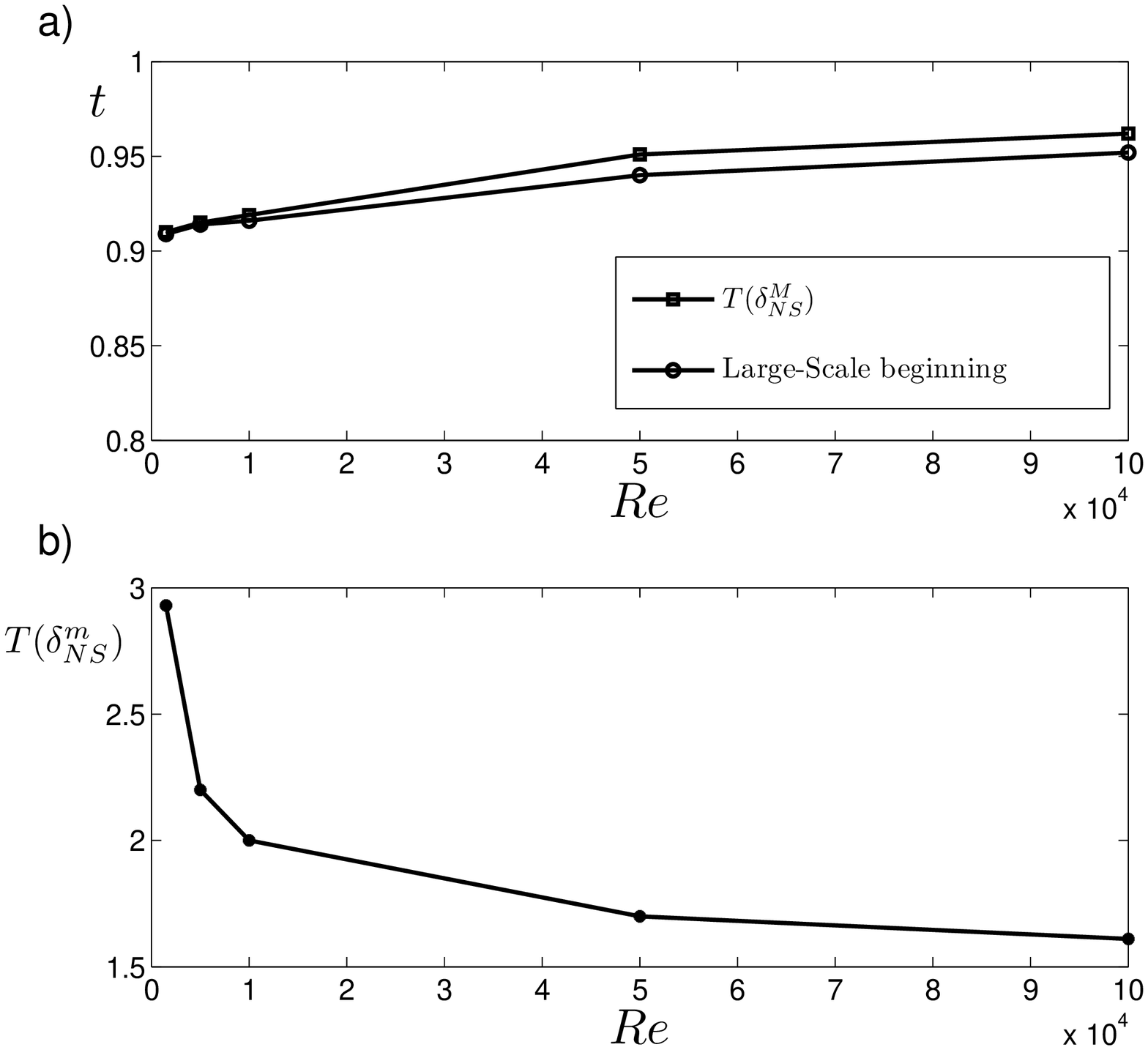}
\caption{\textbf{a)} Comparison of the times at which large-scale
interaction begins ($T_{LS}$) and the time at which 
$\delta_{NS}$ has a local maximum for various Reynolds numbers.  The times agree closely 
and this is a strong indicator that
the beginning of large-scale interaction can be explained in terms of the complex
singularities of the solution.  \textbf{b)} The time $T(\delta_{NS}^m)$ at which
$\delta_{NS}$ has a local minimum in time.  This time gets closer to Prandtl's
singularity time $t_s=1.5$ as Reynolds number increases.}
\label{large_spectrum}
\end{center}
\end{figure}

We begin our analysis by showing in Figure~\ref{delta_NS_inset}a the
temporal evolution of the rate of exponential decay $\delta_{NS}$ evaluated from
the asymptotic behavior of the Fourier amplitudes for various Reynolds numbers. 
This temporal evolution is similar in all 
cases. At $t=0$, the no-slip boundary condition makes $u$ discontinuous on the surface of the circular cylinder
(discontinuity in the normal variable); therefore, $\delta_{NS}(0)=0$.
The effect of viscosity is to regularize the solution and increase the width of the 
analyticity strip $\delta_{NS}$, which reaches 
a maximum value $\delta_{NS}^M$ (see Table~3, in which the time at which $\delta_{NS}(t)$ has its maximum $\delta_{NS}^M$ and its minimum $\delta_{NS}^m$ are reported). 
\begin{table}
\begin{center}
\begin{tabular}{llll}
 \hline
 $\delta_{NS}^{M}-\delta_{NS}^{m}$ &  $Re$ & $T(\delta_{NS}^{M})$ &
$T(\delta_{NS}^{m})$ \\
\hline
$0.332-0.0272$ &$1\cdot10^3$ & $0.908$ & 2.98\\
$0.23-0.01$ &$5\cdot10^3$ & $0.915$ & 2.2\\
$0.21-0.0048$ & $10^4$ & $0.919$ & 2\\
$0.125-0.0011$ & $5\cdot10^4$ & $0.951$ & 1.7\\
$0.096-0.00076$ & $10^5$ & $0.962$ & 1.61\\
\hline
  \end{tabular}
\end{center} 
\caption{The maximum $\delta_{NS}^{M}$ and minimum $\delta_{NS}^{m}$
values of the rate of exponential decay of the Fourier amplitudes, and the times
of their formation ($T(\delta_{NS}^{M})$ and $T(\delta_{NS}^{m})$) for all the Reynolds
numbers considered.}
\label{deltamaxmin}
\end{table}
A comparison with the time of formation of
large-scale interaction $T_{LS}$ (shown in Table~2) shows that the time at which $\delta_{NS}$ has its maximum and the
time at which large-scale interaction begins agree quite well as one can see in
Figure~\ref{large_spectrum}, where the two times are
shown for various Reynolds numbers. In practice after large-scale interaction, the complex singularities of the solution begin to get close to the real domain.
The subsequent decrease of $\delta_{NS}$ is due to the fact that the gradients in the $\theta$ direction become stronger than the gradients in the normal direction $r$. This is 
expected during the large-scale interaction stage when the large gradients of the solution
in the streamwise direction $\theta$, which eventually becomes more pronounced during the small-scale interaction stage, begin to form. 

For all the Reynolds numbers considered, $\delta_{NS}$ has a local minimum $\delta_{NS}^m$ in time
after $t_s=1.5$  (see Table~3), and then begins to
increase.  
A similar behavior in the width of the analyticity strip has been observed in \cite{SSF83}   
for the viscous Burgers equation. In Figure~\ref{delta_NS_inset}b, it is also shown that
$\delta_{NS}^m$ seems to scale linearly with respect to $1/Re$ for the Reynolds
numbers for which small-scale interaction occurs, and as Reynolds number increases.  Moreover, 
the time at which $\delta_{NS}^m$ forms gets very close to Prandtl's singularity time $t_s=1.5$ as shown in
Figure~\ref{large_spectrum}b.  This can be viewed as confirmation of the validity of boundary-layer theory given by the analysis of the complex singularity of the solution of Navier-Stokes,
as for $Re\rightarrow\infty$ we expect that $\delta_{NS}^m\rightarrow0$ at $t_s\rightarrow1.5_{-}$.


Regarding the evaluation of $\alpha_{NS}$, observe that all of the spectra analyzed have several structures, particularly in the 
lower modes where  the algebraic prefactor is generally much easier to capture. 
This renders it very difficult to evaluate $\alpha_{NS}$. 
In order to combat this limitation, a different fitting procedure has been performed, but it also does not give satisfactory results as  
compared to the results for Prandtl's equation by \cite{GSS09}, where the spectrum was straightforward to calculate.
Despite these difficulties, there are some clear signs that allow for conclusions to be drawn. 
First of all, after some time, i.e.\ after the onset of large-scale
interaction, a fitting of the Fourier amplitudes always gives results in the range
$0.45<\alpha_{NS}<0.55$ for all of the Reynolds numbers considered. So it is
very likely that $\alpha_{NS}=1/2$.

The characterization $\alpha_{NS}=1/2$ is compatible with the results 
presented in the previous section for the singularities of
$\tau_w^{NS}$ characterizing the
large- and small-scale interaction. In fact, $s_{ls}$ and $s_{ss}$ are both found to be $1/2$, 
and $\tau_w^{NS}$ is nothing other than the normal derivative of $u$ on the cylinder surface. 
Moreover, close to the circular cylinder $\omega\approx -\partial_r u$. 
Because the gradients of $u$ in the angular (streamwise) direction $\theta$
are greater than those in the normal direction $r$, particularly after the onset of
large-scale interaction, a function whose gradients are still more pronounced along the angular variable is obtained even if $u$  is differentiated with respect to the normal coordinate.  As the relevant gradients of
$u$ and $\partial_r u$ in the angular direction are primarily concentrated close to the cylinder surface, they resemble the behavior of the gradients of $\tau_w^{NS}$, which is simply the rescaled normal derivative of $u$
on the cylinder surface.  Therefore, a match is expected between the characterization of the
complex singularities of $\tau_w^{NS}$ and $u$.  
Recall that in \cite{GSS09}, the characterization of van Dommelen's singularity in Prandtl's equations was $\alpha_{P}=1/3$, 
which is different from that obtained for the Navier-Stokes equations here. 
We suppose that $\alpha_{NS}$ and $\alpha_{P}$ are strongly influenced by the viscous-inviscid interactions occurring during 
the flow evolution at finite Reynolds number.  
We already have seen that the various interactions that occur in Navier-Stokes solutions induce different flow evolution owing to 
the formation of different structures as compared to that forming in Prandtl's case.  
Therefore, a discrepancy between the characterizations of $\alpha_{NS}$ and $\alpha_{P}$ is
likely to occur.

\begin{figure}
\includegraphics[width=12.5cm,]{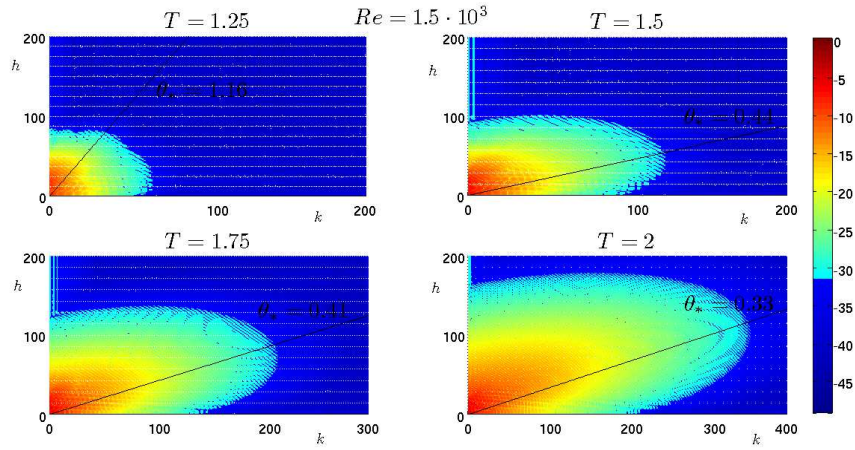}
\caption{The  spectrum of $u$ for $Re=1.5\cdot10^3$ at various
time. }

\label{spetrum2d1500}
\end{figure}
\begin{figure}
\includegraphics[width=12.5cm]{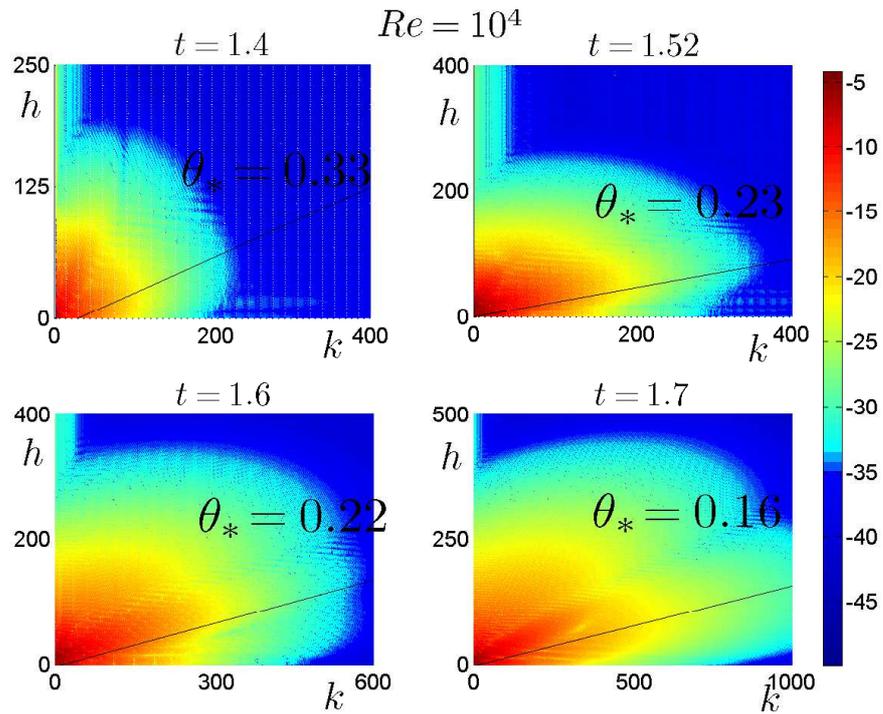}
\caption{The  spectrum of $u$ for $Re=10^4$ at various time. At
$t=1.5$, a bulge forms in the spectrum that becomes more pronounced as time
passes. It can be related to the effect of small-scale interaction.}
\label{spetrum2d10000}
\end{figure}
\begin{figure}
\begin{center}
\includegraphics[width=12.5cm,height=9cm]{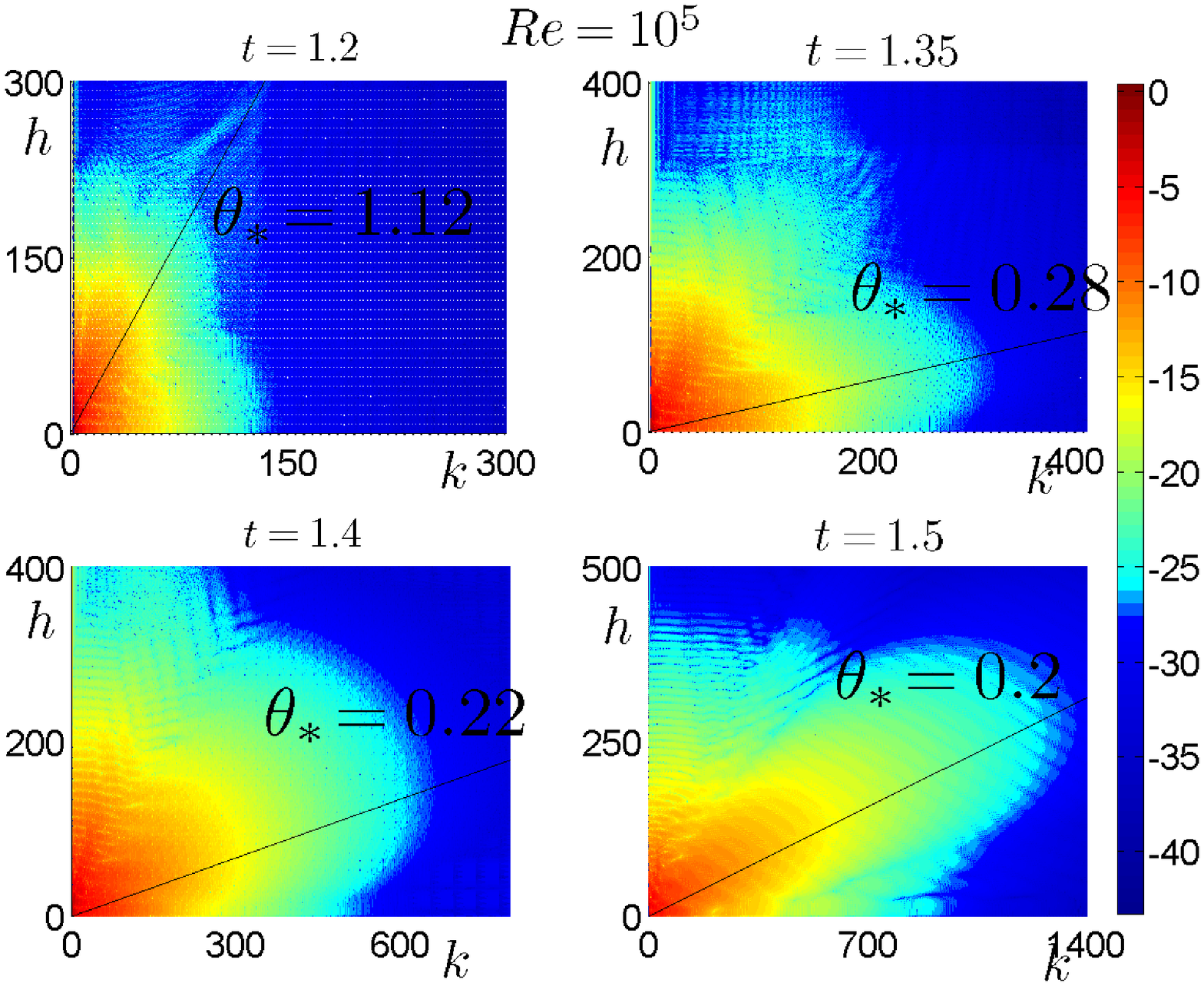}
\caption{The  spectrum of $u$ for $Re=10^5$ at various time. At
$t=1.35$, a bulge forms in the spectrum that becomes more pronounced as time
passes.  The behavior is similar to that shown in Figure~\ref{spetrum2d10000}
for the case $Re=10^4$.}
\label{spetrum2d100000}
\end{center}
\end{figure}

It is also interesting to analyze the most singular direction $\theta^*$ of the
bi-dimensional spectrum of $u$. The spectra are shown in Figures~\ref{spetrum2d1500}, \ref{spetrum2d10000}, and \ref{spetrum2d100000} at different times for $Re=1.5\cdot10^3,10^4,10^5$, and the
most singular direction is indicated by the straight line.
In each case, the rate of exponential decay $\delta(\theta^*)$ of the spectrum along $\theta^*$ exactly coincides with
the value $\delta_{NS}$. 
However, the most relevant attribute is that at a particular time a bulge forms in the
spectrum along the most singular direction.
This is visible only for $Re=10^4,10^5$ (also for
$Re=5\cdot10^3,5\cdot10^4$ not shown here); observe that these are the Reynolds numbers for which small-scale interaction is present, and the time at which the bulge begins to appear matches closely with that at which small-scale interaction begins.
This bulge in the spectrum has a spiky nature  
resembling the physical formation of the kink in the streamlines and vorticity
contours owing to the small-scale interaction. This can be justified on the basis that the small-scale
interaction reveals itself through formation of large gradients in the angular direction
$\theta$ in the solution.  Therefore, it is expected that the spectrum rapidly focuses in 
directions approaching $\theta=0$, where the Fourier modes are much more energized.
It has been observed that as time passes, the most singular direction approaches $\theta^*=0$, which confirms that the 
relevant gradients present in $u$ are those
relative to the coordinate $\theta$. This result is also compatible with
the result predicted by boundary-layer theory.  In fact, in \cite{GSS09} it was
shown that the most singular direction in the spectrum of the streamwise velocity
component $u$ of Prandtl's equation at the singularity time
$t_s=1.5$ is $\theta^*=0$, meaning that the blow up occurs
in $\partial_x u$ as reported previously in Section \ref{PS}.
This behavior is not visible for the case $Re=10^3$ (see
Figure~\ref{spetrum2d1500}). In fact, the spectrum grows throughout a wider range
around the most singular direction, and the bulge in the spectrum is not
present because no large gradients form in the solution.

\section{Conclusions}
Solutions of the two-dimensional Prandtl and Navier-Stokes equations have been obtained for
the  case of flow past an impulsively-started circular cylinder using a fully spectral numerical scheme in which the
solution is approximated by Fourier-Chebyshev expansions.
Based on these solutions, the asymptotic validity of unsteady boundary-layer theory has been analyzed by comparing
Prandtl's solution with Navier-Stokes solutions at different Reynolds numbers in the range $Re=10^3$ to $Re=10^5$.
Prandtl's equations develop a singularity at the finite time $t_s=1.5$ owing to the lack of interaction between the viscous boundary layer and inviscid outer flow, which is anticipated at $t_k\approx1.4$. This interaction can be detected in the Navier-Stokes solutions by the formation of a kink in the streamlines and the formation of a local
maximum in the displacement thickness.

The unsteady separation process for Navier-Stokes solutions reveals a different behavior.  In particular, the presence 
of viscous-inviscid interactions acting over different scales can be detected.
The first interaction is found to occur for all Reynolds numbers considered, and it
is marked by discrepancies arising between the streamwise pressure gradient and wall shear on the surface of the circular cylinder and the same quantities
predicted by Prandtl's equations. In particular, we have defined the
beginning of large-scale interaction to be formation of an inflection point in the streamwise
pressure gradient on the cylinder surface close to its local maximum.  This represents a
different topological structure as compared to the streamwise pressure gradient
predicted by Prandtl's equation.

The large-scale interaction is distinct from the viscous-inviscid interaction that develops in the infinite Reynolds number case, i.e.\
prior to formation of the van Dommelen singularity in Prandtl's solution, as no
large gradients or spiky behavior are visible in Navier-Stokes solution for such cases. However, this
interaction is the precursor to the small-scale interaction that occurs at higher Reynolds numbers and is 
similar to the interaction that occurs in the infinite Reynolds number case. In fact, small-scale interaction,
which occurs only for moderate to high Reynolds numbers ($Re\geq O(10^4$)), is
characterized by formation of a kink in the streamlines and vorticity
contours. This stage is followed by splitting of the primary recirculation region,
formation of large gradients in the wall shear and streamwise pressure gradient,
and formation of dipolar structures within the boundary layer that cause the
production of large amounts of vorticity and a corresponding growth of the enstrophy. The
beginning of small-scale interaction has been defined by the formation of a local
minimum in the streamwise pressure gradient on the circular cylinder, which plays a central role in
formation of the unsteady separation phenomena.

The various interactions occurring in Prandtl and Navier-Stokes solutions have been
investigated by performing a complex singularity analysis on the wall shear stress and on the velocity component
$u(r,\theta)$ from Navier-Stokes solutions .
To investigate the complex singularities of the wall shear, we have used the
singularity-tracking method, the BPH method, and a method based on Pad\'{e} approximations. It has been found that Prandtl's wall shear becomes singular at $t_s=1.5$ as a
singularity having $\alpha\approx7/6$ hits the real axis near $x_s\approx1.94$.  The 
Navier-Stokes wall shear shows a more complicated behavior during the various stages of
separations that is related to the presence of three different groups of complex
singularities. The first group is formed only by the singularity $s_P$, which
has the same characteristics as van Dommelen's singularity in Prandtl's equations.  This singularity can
be related physically to the birth of the recirculation region. At the 
singularity time $t_s=1.5$, the distance $y_{P}$ of $s_P$ from the real axis
follows the relationship $y_{P}=C_{P}Re^{\lambda_P}$, with $\lambda_P\approx-0.25$ and $C_P\approx3.2$.
The second group of singularities is connected to formation of the large-scale interaction that occurs for all finite Reynolds numbers.
It is the differences that arise between the Prandtl and Navier-Stokes wall shear that are the indicator
of the onset of large-scale interaction. These differences are due to the
formation of a gradient in Navier-Stokes wall shear that corresponds to a complex
singularity $s_{ls}$ having the characteristic value $\alpha = 0.5$. This singularity is quite
well characterized through the singularity-tracking method, while the other
singularities of the group can only be observed by evaluating the Pad\'{e} approximation of the wall shear. 
At the time in which large-scale interaction forms, the singularity $s_{ls}$ moves closer to the real
axis as Reynolds number increases, and it has been found that its distance $y_{ls}$ from the real axis
follows the relationship $y_{ls}=C_{ls}Re^{\lambda_{ls}}$, where 
$\lambda_{ls}\approx-0.138$ and $C_{ls}\approx0.447$.

The small-scale interaction, which appears only for moderate to high Reynolds numbers, 
is characterized by the formation of large gradients in the wall
shear near the cylinder surface beneath the kink in the recirculation region.
These gradients are due to the presence of a third group of complex
singularities, in which the relevant singularity $s_{ss}$ is also
characterized by the value $\alpha = 0.5$. At the time at which small-scale interaction begins, the
distance $y_{ss}$ of $s_{ss}$ from the real axis follows the relationship 
$y_{ss}=C_{ss}Re^{\lambda_{ss}}$, where $\lambda_{ss}\approx-0.25$ and
$C_{ss}\approx0.41$.  In addition, all of the complex singularities of
$\tau_w^{NS}$ converge to one another, i.e.\ focusing in a smaller region, as Reynolds number increases,
suggesting that asymptotically all of the singularities collapse into only a single 
singularity, which is van Dommelen's singularity $s_P$, and all the viscous-inviscid 
interactions reduce to the only interaction observed in the limiting Reynolds number case.

The singularity-tracking method has also been applied to the shell summed
amplitudes of the Fourier-Chebyshev coefficients of the streamwise velocity
component $u(r,\theta)$.
The width $\delta_{NS}$ of the analyticity strip of the
Navier-Stokes solutions for various Reynolds numbers has been tracked in time. The initial discontinuity of $u(r,\theta)$ on
the cylinder surface owing to the no-slip condition makes $\delta_{NS}=0$ at $t=0$. The regularizing
effect increases $\delta_{NS}$, and a maximum value of $\delta_{NS}^M$ forms at
the beginning of the large-scale interaction for all Reynolds numbers considered.  At
this time, large gradients form in the solution in the $\theta$ direction, and they become stronger during the small-scale interaction. Therefore, $\delta_{NS}$ decreases in time as the complex singularities that originate these
gradients get close to the real axis.  The subsequent temporal evolution of
$\delta_{NS}$ reveals an interesting aspect as $\delta_{NS}$
continues to decrease, reaching a minimum value $\delta_{NS}^m$, in time and
then begins to increases again.  The minimum $\delta_{NS}^m$ is $O(1/Re)$ for the Reynolds numbers for which
small-scale interaction forms, and it forms after van Dommelen's singularity time $t_s=1.5$.  However, as Reynolds number
increases, the time of formation of $\delta_{NS}^m$ tends to $t_s=1.5$,
supporting the validity of the predictions from 
boundary-layer theory according to which $\delta_{NS}^m=0$ at $t_s=1.5$ as
$Re\rightarrow\infty$. Moreover, boundary-layer theory predicts that at the singularity time, 
the most singular direction in the bi-dimensional spectrum is $\theta^*=0$ with formation of a
singularity in the streamwise direction.  In the Navier-Stokes solutions, the most
singular direction in the spectrum is indeed found to be close to $\theta^*=0$ as Reynolds number increases, particularly after large-scale interaction begins. From the analysis of the bi-dimensional spectrum of $u(r,\theta)$, the small-scale interaction can also be characterized because, during this stage, the spectrum focuses in a
very narrow zone along the most singular direction, forming a bulge that in some
sense resembles the physical formation of the kink and spike in the streamlines. The
primary difference between the analysis of the spectrum of $u(r,\theta)$ as compared to that from Prandtl's solution is the characterization obtained from the
rate of algebraic decay of the shell summed amplitudes.  For Navier-Stokes solutions, for example, 
it has been found that $\alpha^{NS}\approx1/2$ for all Reynolds numbers considered, while
the prediction of boundary-layer theory is that $\alpha^P\approx1/3$.  This discrepancy 
can be explained by the presence of the large- and small-scale interactions
that act in a different manner on the flow evolution as compared to the viscous-inviscid 
interaction present as $Re\rightarrow\infty$.  In this limit, the large- and small-scale interactions eventually merge together to become the
viscous-inviscid interaction observed in the infinite Reynolds number case, which should
lead to alignment of $\alpha^{NS}$ to the value $1/3$.

Even though the overall results of this paper strongly suggest that the finite-Reynolds number Navier-Stokes
viscous-inviscid interactions tend to that predicted by boundary-layer
theory in the limit as Reynolds number goes to infinity, one should keep in mind the
possible presence of a Rayleigh instability, which has been observed in solutions of the Navier-Stokes equation for very high Reynolds numbers in other geometries. For example, \cite{CO10}  show that for
the thick-core vortex, an instability develops in the form of high-frequency
oscillations in vorticity and streamwise pressure gradient along the wall in the
range $Re=10^5-10^8$. This instability is of Rayleigh-type, with the dominant
wavenumber being $O(Re^{1/2})$, meaning that the disturbances are
of the same streamwise scale as the boundary-layer thickness, which is $O(Re^{-1/2})$. 

These results are somehow
related to that obtained by \cite{GR00}, in which a class of initial
profiles for which Euler equations are linearly
unstable due to an exponential growth of modes of size $Re^{1/2}$, i.e.\ a phenomenon linked to Rayleigh instability, does not allow the solution to have the form of a matched asymptotic expansion
between a Prandtl solution and an Euler solution.

The appearance of a Rayleigh instability can be detected only by using a
very refined grid in the zone where instability forms (see \cite{CO10}) in order to avoid
any numerical disturbances that can create possible doubts in the nature of the
instability.  For example, \cite{BW02} observed an instability acting in a similar fashion
as observed in \cite{CO10} and attributed it to a Rayleigh instability; however, the oscillations were found to disappear when 
using a finer grid in \cite{OC05}. We have also checked to see if 
a Rayleigh instability occurs in the impulsively-started circular cylinder case for very high Reynolds numbers
($Re=10^6,10^7$). Even if spurious oscillation appeared in the solution in the same fashion
as observed for the thick-core vortex, we were not able to discern between numerical
or physical instability owing to a lack of the required numerical
resolution.  We plan to investigate the physical instabilities that may occur in Navier-Stokes solutions for this
initial datum with the aid of a more powerful computing resource.
Moreover, it should be very interesting to search for a possible link between the
Rayleigh instability and the presence of complex singularities in the solutions, and
this topic will be the object of our future work.

\bibliographystyle{amsalpha}
\bibliography{prandtl}

\end{document}